\documentclass[11pt]{article}
\usepackage{amssymb,latexsym,amsmath, amsthm}
\usepackage{amsfonts, graphics,color}
\usepackage{mathtools}
\usepackage{hyperref} %Für Links
\usepackage{breakurl} %Fuer Linebreak beim Link (beim uebersetzen vom arxiv) 
\usepackage[american]{babel}
\usepackage{csquotes}

\usepackage{subcaption} %subfigures
\usepackage{float}

\def\R{\mathbb R}

\let\originalleft\left
\let\originalright\right
\renewcommand{\left}{\mathopen{}\mathclose\bgroup\originalleft}
\renewcommand{\right}{\aftergroup\egroup\originalright}
\newcommand{\ft}[0]{\footnotesize}

\theoremstyle{remark}

\title{A numerical stability analysis for the Einstein-Vlasov system}
\author{Sebastian G\"unther\thanks{Department of Mathematics, University of Bayreuth, Germany}~, Jacob K\"orner\thanks{Institute of Mathematics, Julius-Maximilians-Universit\"at W\"urzburg, Germany}~, Timo Lebeda\thanks{Department of Physics, University of Bayreuth, Germany}~, Bastian P\"otzl\footnotemark[1]~,\\
Gerhard Rein\footnotemark[1]~, Christopher Straub\footnotemark[1]~, J\"org Weber\thanks{Centre for Mathematical Sciences, Lund University, Sweden}}

\begin{document}
\maketitle

\begin{abstract} 
  We investigate stability issues for steady states of the spherically symmetric
  Einstein-Vlasov system numerically in Schwarzschild, maximal areal, and
  Eddington-Finkelstein coordinates. Across all coordinate systems we confirm
  the conjecture that the first binding energy maximum along a one-parameter
  family of steady states signals the onset of instability. Beyond this
  maximum perturbed solutions either collapse to a black hole, form
  heteroclinic orbits, or eventually fully disperse. Contrary to earlier
  research, we find that a negative binding energy does not necessarily
  correspond to fully dispersing solutions. We also comment on the so-called
  turning point principle from the viewpoint of our numerical results.
  The physical reliability of the latter is strengthened by obtaining consistent
  results in the three different coordinate systems and by
  the systematic use of dynamically accessible
  perturbations.  
\end{abstract}

\maketitle
\section{Introduction}
\setcounter{equation}{0}
We consider in the context of general relativity
a large ensemble of mass points which interact only through
the gravitational field which they create collectively.
Such a self-gravitating collisionless gas is used in astrophysics
to model galaxies or globular clusters.
Gravity is described by the Einstein equations
\begin{equation} \label{feqgen}
G_{\alpha \beta} = 8 \pi T_{\alpha \beta},
\end{equation}
where $G_{\alpha \beta}$ is the Einstein tensor induced by the
Lorentzian metric $g_{\alpha \beta}$
with signature $(-{}+{}+{}+)$ on the smooth spacetime manifold $M$, and
$T_{\alpha \beta}$ is the energy-momentum tensor given by the
matter content of the spacetime. Greek indices run from $0$ to $3$,
and we choose units in which the speed
of light and the gravitational constant are equal to $1$.
The evolution equation for a collisionless gas is
the collisionless Boltzmann or Vlasov equation
so that we obtain the Einstein-Vlasov system.
We study this system under the assumption that the
spacetime is spherically symmetric and asymptotically flat,
but we first formulate it in general.

The world line of a test particle
on $M$ obeys the geodesic equation
\[
\dot x^\alpha = p^\alpha,\ 
\dot p^\alpha = - \Gamma^\alpha_{\beta \gamma} p^\beta p^\gamma,
\]
where $x^\alpha$ denote general coordinates on $M$, 
$p^\alpha$ are the corresponding canonical momenta,
$\Gamma^\alpha_{\beta \gamma}$ are the Christoffel symbols
induced by the metric $g_{\alpha \beta}$,
the dot indicates differentiation with respect to proper time along the
world line of the particle, and the Einstein summation convention is applied.
We assume that all the particles in the ensemble have the same
rest mass, normalized to $1$, and move forward in time, i.e.,
their number density $f$ is a non-negative function
supported on the mass shell
\[
PM := \left\{ g_{\alpha \beta} p^\alpha p^\beta = -1,\ p^\alpha \
\mbox{future pointing} \right\},
\]
a submanifold of the tangent bundle $TM$ of the spacetime manifold $M$
which is invariant under the geodesic flow.
Letting Latin indices range from $1$ to $3$
we choose coordinates $(t,x^a)$ such that on the mass shell
$PM$ the variable $p^0$ becomes 
a function of the remaining variables $(t,x^a,p^b)$; $t$ should be
thought of as a time-like variable.
Since the particles in the ensemble move like test particles,
their number density
$f=f(t,x^a,p^b)$ is constant along the geodesics and hence satisfies
the Vlasov equation
\begin{equation} \label{vlgen}
\partial_t f + \frac{p^a}{p^0}\,\partial_{x^a} f
-\frac{1}{p^0}\,\Gamma^a_{\beta \gamma} p^\beta p^\gamma\,\partial_{p^a} f = 0.
\end{equation}
The energy-momentum tensor is given by
\begin{equation} \label{emtvlgen}
T_{\alpha \beta}
=\int p_\alpha p_\beta f \,|g|^{1/2} \,\frac{dp^1 dp^2 dp^3}{-p_0},
\end{equation}
where $|g|$ denotes the modulus of the determinant of the metric,
and indices are raised and lowered using the metric, i.e.,
$p_\alpha = g_{\alpha \beta}p^\beta$.
The system \eqref{feqgen}, \eqref{vlgen}, \eqref{emtvlgen}
is the Einstein-Vlasov system in general coordinates. We
want to model isolated systems and therefore require that the
spacetime is asymptotically flat. In order to simplify the system
we only consider spherically symmetric solutions. In the next section we
formulate the Einstein-Vlasov system in coordinates adapted to this
symmetry. For background on the Einstein-Vlasov system we refer
to \cite{An2011} and the references there.

The Einstein-Vlasov system possesses a plethora of steady state solutions.
For a stationary metric the Killing vector $\partial/\partial t$
gives rise to the quantity $E = -g(\partial/\partial t,p^\alpha)$ which
represents the particle energy  and is constant along geodesics.
Hence the ansatz
\begin{equation} \label{miceqstate}
f(x^a, p^b) = \phi(E)
\end{equation}
satisfies the stationary Vlasov equation and reduces the system to the
field equations. In Section~\ref{ssc:steady} we recall how any such
\enquote{microscopic equation of state $\phi$} gives rise to a one-parameter
family of steady states, where the parameter can be identified with the central
redshift of the configuration and is therefore a measure of how relativistic
it is; the steady states actually considered below can also depend on the angular momentum of the particles. A natural question is which of these steady states are stable or
unstable. For the Vlasov-Poisson system, which is the non-relativistic limit
of the Einstein-Vlasov system, such steady states essentially are stable if
the microscopic equation of state $\phi$ is a decreasing function of the
particle energy, cf.\ \cite{Rein2007} and the references there.
For the relativistic case of the Einstein-Vlasov system the situation is
quite different. At least on the linearized level it has been shown in
\cite{HaRe2013,HaRe2014,HaLiRe2020} that such steady states are stable if their
central redshift is sufficiently small, but for the same microscopic
equation of state they become unstable if their central redshift is large.
The question whether sufficiently relativistic matter distributions become
unstable played an important role in the discovery and subsequent
discussion of quasars, cf.\ \cite{BKZ,IT68,Ze1971}. But
unstable steady states of the Einstein-Vlasov system are also important
for conceptual reasons, since they can possibly explain the so-called
type~I behavior in critical collapse observed for the Einstein-Vlasov system,
the latter being related to the cosmic censorship hypothesis,
cf.\ \cite{AnRe2006,OC,ReReSch}. All this motivates the present investigation
where we analyze the transition from stability to instability along
one-parameter families of steady states of the Einstein-Vlasov system by
numerical means. In particular, we investigate where this transition takes place
and what happens to weakly perturbed steady states which lie in the unstable regime.

Concerning the former question there are various possibilities.
One can for example consider the so-called binding energy as a function of the central redshift, cf.~Figure~\ref{img:bindeng}.
It has been conjectured \cite{Ze1966, Ze1971} that the transition from stability to instability
happens at the first (local) maximum of this curve.
For all microscopic equations of state which we consider we confirm this conjecture.

Alternatively, one can plot for a fixed microscopic equation of state
and each value of the central redshift the ADM mass and radius of the support
of the corresponding steady state. This results in a so-called mass-radius
curve, cf.~Figure~\ref{img:spirale}. It would be conceivable that the stability properties change at the
turning points of this curve since a precise version of this so-called turning point
principle has recently been proven both for the Euler-Poisson and
the Einstein-Euler system, cf. \cite{HaLi, LiZe}. In these models, matter is described as an
ideal, compressible fluid. The macroscopic quantities induced by an
isotropic steady state of the Einstein-Vlasov system of the form
\eqref{miceqstate} yield a steady state of the Einstein-Euler system with a suitable, induced
macroscopic equation of state,
cf. \cite{HaLiRe2020}. In particular, the mass-radius curves are then
the same for both systems. However, we clearly disprove this turning point principle for the Einstein-Vlasov system.

Another issue is to understand the behavior of solutions which are
launched by small perturbations of a stable or an unstable steady state.
In the former case we find that the system starts to oscillate, i.e.,
to expand and contract in a seemingly time-periodic fashion. This behavior
was observed in \cite{RaRe2013} for the Vlasov-Poisson system
and in \cite{AnRe2006} for shell-like solutions of the Einstein-Vlasov
system; the code employed in \cite{AnRe2006} was not able to properly
handle steady states which have matter at the center instead of a
vacuum region. The behavior of an unstable steady state after perturbation
is more interesting. The solution either collapses and forms a black hole, or it seems
to follow a heteroclinic orbit to a different, stable steady state
about which (the bulk of) it starts to oscillate. The terminology
\enquote{heteroclinic orbit} may not be quite appropriate here, but it captures
the observed behavior. In \cite{ShTe1985} a similar observation is claimed without further comment.
For steady states with a large central redshift the perturbed state may also disperse instead of following a
heteroclinic orbit as explained above.

In all the numerical simulations we used dynamically accessible perturbations
which in particular preserve all the so-called Casimir functionals \eqref{eq:casimir}
of the system; perturbing the steady state by some external force
results in such dynamically accessible states.
All the simulations were performed in three different coordinate systems,
namely Schwarzschild, maximal areal, and
Eddington-Finkelstein coordinates, and our observations were completely
consistent across these. This is a priori not obvious. In a stability
analysis for the Einstein-Vlasov system one necessarily must compare
functions like metric components or mass-energy densities which are
defined on two different spacetimes, the stationary one and the perturbed one.
There is no canonical way of identifying points on these two spacetimes
so that one could compare the values of certain functions at
those identified points. What we do is to simply identify points which have
the same coordinates in the coordinate system at hand, and it is therefore
not a priori clear that the stability findings in different coordinate systems
must be consistent. This point has also been made in the
astrophysics literature, cf.\ \cite{IT68}.

The paper proceeds as follows. In the next section we formulate
the Einstein-Vlasov system in the coordinate systems mentioned above,
compare these coordinate systems, and recall how steady states of the
Einstein-Vlasov system are obtained. In Section~\ref{sc:numeric}
we explain the numerical method which we employ, which is a particle-in-cell
scheme and lends itself well to parallelization.
The numerical results are presented and discussed in Section~\ref{sc:results},
and in the last section we comment on their precision and reliability.
\section{The spherically symmetric Einstein-Vlasov system}
\setcounter{equation}{0}
In this section we formulate the Einstein-Vlasov system in Schwarzschild,
maximal areal, and Eddington-Finkelstein coordinates and compare various
properties of these coordinate systems.
We also recall how steady state solutions can be obtained. 

\subsection{Schwarzschild coordinates}\label{ssc:ss}
In Schwarzschild coordinates the metric reads
\begin{align} 
  ds^2 = -e^{2\mu(t,r)} dt^2 + e^{2\lambda(t,r)} dr^2
  + r^2 \left( d\theta^2 + \sin^2\theta d\varphi^2 \right) \label{eq:metric}
\end{align}
with
$\left( t, r, \theta, \varphi \right) \in
\R \times \left[ 0, \infty \right[ \times
\left[ 0, \pi \right] \times\left[ 0, 2\pi \right]$
and metric coefficients $\mu=\mu(t,r)$ and $\lambda=\lambda(t,r)$.
Here, $t$ corresponds to the proper time of an observer located at
spatial infinity and $r$ denotes the areal radius. For numerical and
analytical reasons it is convenient to introduce Cartesian coordinates 
\[
x =(x^1, x^2, x^3)=
r(\sin\theta \cos\varphi, \sin \theta \sin \varphi , \cos \theta ) \in \R^3
\]
and corresponding non-canonical momentum variables
\[
v^i = p^i + \left ( e^{2\lambda} -1 \right ) \frac{x \cdot p}{r} \frac{x^i}{r} .
\]
Since we consider the asymptotically flat case and in order to guarantee
a regular center, we impose the boundary conditions
\begin{align} \label{eq:boundary_conditions}
  \lim_{r \to \infty} \lambda \left( t, r \right) =
  \lim_{r \to \infty} \mu \left( t, r \right) = \lambda \left( t, 0 \right) = 0,
  \ t \in \R.
\end{align}
Inserting the metric into the Einstein equations yields the following
field equations:
\begin{align}
  e^{-2\lambda} \left( 2r\lambda' - 1 \right) +1
  &= 8 \pi r^2 \rho, \label{eq:EV_rho}
\\
e^{-2\lambda} \left( 2r\mu' + 1 \right) -1
&= 8 \pi r^2 p, \label{eq:EV_p}
\\
\dot{\lambda} = -4 \pi
&r e^{\mu + \lambda} j. \label{eq:EV_j}
\end{align}
These are the $00$, $11$, and $01$ components of the general equation
\eqref{feqgen}.
Equation \eqref{eq:EV_j} is not independent, but follows from
\eqref{eq:EV_rho} and \eqref{eq:EV_p} together with the Vlasov equation.
However, it is useful for the numerics. The also non-trivial
$22$ and $33$ components of \eqref{feqgen}
follow as well, but they are not used in the numerics.
In the above, $\dot{}$ and ${}'$ denote the derivative with respect to
$t$ or $r$ respectively.
The Vlasov equation takes the form
\begin{align}
\partial_t f + e^{\mu - \lambda} \frac{v}{\varepsilon} \cdot \partial_x f
- \left(\dot{\lambda} \frac{x\cdot v}r + \mu' e^{\mu - \lambda} \varepsilon \right)
\frac{x}{r} \cdot \partial_v f = 0, \label{eq:EV_Vlasov}
\end{align}
where we introduce
\begin{align*} %\label{eq: vareps(r,w,L)}
\varepsilon = \sqrt{ 1 + \vert v\vert^2} = \sqrt{ 1 + w^2 + \frac{L}{r^2}} .
\end{align*}
Here $|v|$ denotes the Euclidean length and $x\cdot v$ the Euclidean scalar
product.
The variables $w = \frac{x\cdot v}{r}$ and $L=\vert x \times v\vert ^2$ can be
thought of as the momentum in the radial direction and the square of the angular
momentum respectively. We assume $f$ to be spherically symmetric, i.e.,
$f(t,x,v)=f(t,Ax,Av)$ for $A \in \mathrm{SO}(3)$, and 
under abuse of notation we may write
$f(t,x,v)=f(t,r,w,L)$.
The source terms in the field equations are defined by
\begin{align}
  \rho(t,r)
  &= \frac{\pi}{r^2} \int_0^\infty \int_{-\infty}^\infty \varepsilon f(t,r,w,L)
  \, dw\, dL, \label{eq:rho(r,w,L)}
  \\ \nonumber \\
  p(t,r)
  &= \frac{\pi}{r^2} \int_0^\infty \int_{-\infty}^\infty \frac{w^2}{\varepsilon}
  f(t,r,w,L) \, dw\, dL, \label{eq:p(r,w,L)}
  \\ \nonumber \\
  j(t,r)
  &= \frac{\pi}{r^2} \int_0^\infty \int_{-\infty}^\infty w f(t,r,w,L) \, dw \,dL.
  \label{eq:j(r,w,L)}
\end{align}
Here, $\rho$ is the energy density, $p$ the radial pressure, and $j$ the
particle current. Equations
\eqref{eq:boundary_conditions}--\eqref{eq:j(r,w,L)}
constitute the Einstein-Vlasov system in Schwarzschild coordinates.
Unless stated otherwise, we employ the above notation for the other
coordinate systems as well.

For later use, we derive some formulas for the metric coefficients.
To this end, we introduce the Hawking mass defined as
\begin{align} \label{eq:def_m(r)_Hawking mass}
m(t,r) = 4 \pi \int_0^r  \rho(t,s) s^2  ds.
\end{align}
Integrating \eqref{eq:EV_rho} yields
\begin{align} \label{eq:exp(-2lambda)}
e^{-2\lambda} = 1 - \frac{2m}{r};
\end{align}
the right hand side of this equation remains positive as long
as the solution to \eqref{eq:boundary_conditions}--\eqref{eq:j(r,w,L)}
exists.
Solving the field equation \eqref{eq:EV_p} for $\mu'$ and
using \eqref{eq:exp(-2lambda)}, we obtain
\begin{align} \label{eq:mu'}
\mu' = e^{2\lambda} \left( 4 \pi r p + \frac{m}{r^2} \right).
\end{align}
\subsection{Maximal areal coordinates}\label{ssc:ma}
In maximal areal coordinates the
line element can be written as
\[
ds^2 = (-\alpha^2 + a^2 \beta^2) dt^2 + 2a^2\beta dt dr + a^2 dr^2 +
r^2( d\theta^2 + \sin^2\theta d\varphi^2) 
\]
with positive metric coefficients $a$ and $\alpha$.
We note that in the present case $t$ is not introduced as
the proper time of any physical observer, but is fixed
by imposing the maximal gauge condition, i.e.,
each hypersurface of constant $t$ has vanishing mean curvature.
The non-canonical momentum variables introduced in the Schwarzschild
case translate to
\[
v_i = p_i + \left ( \frac1a -1 \right ) \frac{x \cdot p}{r} \frac{x_i}{r} .
\]
In analogy to \eqref{eq:boundary_conditions} the metric coefficients
satisfy the boundary conditions
\begin{align} \label{eq:boundaryma} 
  a(t,0) = \lim_{r\to\infty} a(t,r) = \lim_{r\to\infty} \alpha(t,r) = 1, \quad
  \beta(t,0) = 0.
\end{align} 
We obtain the field equations 
\begin{align}
  \kappa
  &= \frac{\beta}{\alpha r} \label{eq:0}, \\
  a'
  &= 4\pi r \rho a^3 + \frac{3}{2} r \kappa^2 a^3  +
  \frac{a}{2r} \left(1-a^2\right)  \label{eq:i}, \\
  \kappa'
  &= - 3\frac{\kappa}{r} -  4\pi a j  \label{eq:ii},\\
  \alpha ''
  &= \alpha ' \left ( \frac{a'}{a} -\frac2r \right ) +
  6\alpha a^2 \kappa^2 + 4\pi \alpha a^2 (S +\rho) \label{eq:iv}, 
\end{align}
which are coupled to the Vlasov equation
\begin{align}\label{eq:vleqfull}
  \partial_t f +
  \left [ \frac{\alpha}{a} \frac v \varepsilon -\beta \frac xr \right ]
  \cdot \partial_x f +
  \left [ - \varepsilon \frac{\alpha'}{a} \,  \frac xr +
    \alpha \kappa \left ( v - 3 \frac{x\cdot v}r \frac xr \right ) \right ]
  \cdot \partial_v f = 0   
\end{align}
via the source terms \eqref{eq:rho(r,w,L)}, \eqref{eq:j(r,w,L)}, and
\begin{align}
  S(t,r)
  &= \frac\pi{r^2} \int_0^\infty \int_{-\infty}^\infty \frac{\varepsilon^2-1}
  \varepsilon  f(t,r,w,L) \, dw \,dL,
\end{align}
the trace of the spatial part of the energy-momentum tensor.

As in Schwarzschild coordinates we introduce the Hawking mass,
which in maximal areal coordinates becomes 
\[ 
m = \frac r 2 \left (1 - \frac{1}{a^2} + r^2 \kappa^2 \right ).
\]
In order to solve the field equations numerically,
it is useful to consider the quantity 
\[
\eta = \frac{r}{2} \left ( 1-  \frac{1}{a^2} \right ),
\] 
which immediately implies
\begin{equation}\label{eq:mass_ma} 
  \eta (t,r) = \int_0^r \left ( 4\pi  \rho (t,s) + \frac 3 2  \kappa^2(t,s)
  \right ) s^2  ds.
\end{equation}
Furthermore, the field equation \eqref{eq:ii} yields the implicit formula
\[ 
\kappa(t,r) = - \frac{4\pi}{r^3} \int_0^r a(t,s) j(t,s) s^3  ds ,
\]
while 
\[ 
\alpha'(t,r) = \frac{a(t,r)}{r^2}
\int_0^r \left ( 4\pi a \alpha (\rho+S) + 6 a \alpha \kappa^2 \right ) s^2  ds
\]
holds because of the second order equation \eqref{eq:iv}. 
Note that $\kappa(t,r) \sim r^{-3}$ for large $r$ provided
that the matter is compactly supported. 
\subsection{Eddington-Finkelstein coordinates} \label{ssc:ef} 
In Eddington-Finkelstein coordinates the metric takes the form
\begin{align*}
  ds^2= -a(t,r)b^2(t,r) dt^2  +2b(t,r) dt dr +
  r^2 (d\theta^2 + \sin^2 \theta \, d \varphi^2).
\end{align*}
Similar to \eqref{eq:boundary_conditions} and \eqref{eq:boundaryma}
the metric coefficients $a$ and $b$ satisfy the boundary conditions
\begin{align}\label{eq:boundinfEF}
a(t,0) = \lim_{r \to \infty} a(t,r) = \lim_{r\to \infty} b(t,r) = 1.
\end{align}
Notice that the metric coefficient $a$ here is not the same as the
coefficient $a$ appearing in maximal areal coordinates, but we
nevertheless use this quite common, albeit equivocal notation.
As opposed to Schwarzschild and maximal areal coordinates we
use the canonical momentum coordinates $(p_0, p_1, p_2, p_3)$.
The angular momentum is given by
\begin{align*}
L = (p_2)^2 + \frac{1}{\sin^2 \theta} (p_3)^2.
\end{align*}
The particle density $f$ can be written as a function of
$(t,r,p_1,L)$ and the Vlasov equation reads 
\begin{align*}
  \partial_t f &+ \frac{b}{2} \Big( a - \frac{1+L/r^2}{(p_1)^2} \Big)
  \partial_r f 
\\
&+ \frac{1}{2} \Bigg( \frac{2bL}{r^3 p_1}-\partial_r (ab)p_1 -
\partial_r b \frac{1+L/r^2}{p_1} \Bigg) \partial_{p_1}f =0.
\end{align*}
Here, the Hawking mass $m$ is given by
\begin{align*}
m = \frac{r}{2}(1-a).
\end{align*}
The metric coefficients $a$ and $b$ as well as the
Hawking mass can be computed directly from $f$ via
\begin{align*}
b(t,r)&=\exp\left(-4\pi\int_r^\infty\eta
T_{11}(t,\eta)\,d\eta\right),\\
m(t,r)&=\frac{2\pi}{b(t,r)}\int_0^r\eta^2(T_{11}+S)(t,\eta)b(t,\eta)\,d\eta,
\nonumber\\
a(t,r)&=1-\frac{2m(t,r)}{r},
\end{align*}
where
\begin{align*}
T_{11}(t,r)&=\frac\pi{r^2}\int_0^\infty\int_0^\infty
p_1f(t,r,p_1,L)\,dL\,dp_1,\\
S(t,r)&=\frac\pi{r^2}\int_0^\infty\int_0^\infty\frac{1+\frac
L{r^2}}{p_1}f(t,r,p_1,L)\,dL\,dp_1.
\end{align*}
We have to keep in
mind that the physical interpretation of the timelike variable $t$
differs across the three coordinate systems, and we should mention
that in Eddington-Finkelstein coordinates, what we called $t$ is
usually denoted as $v$.
\subsection{Properties and comparison of the coordinate systems}
\label{ssc:comparison}
Before we investigate the system numerically, we briefly discuss and
compare selected properties of the coordinate systems. Two conserved
quantities of the system are the total number of particles
or total rest mass and the ADM mass. 
Across all coordinate systems the latter is given by
\begin{align}
M = \lim_{r \to \infty} m(t,r) .
\end{align}
For compactly supported matter this equals the Hawking mass
evaluated at the outer boundary of the radial support. 

The total number of particles is computed differently across the
coordinate systems. In the Schwarzschild case we have
\begin{align} \label{eq:def_Nss(t)}
  N = 4 \pi^2 \int_0^\infty \int_{-\infty}^\infty \int_0^\infty e^{\lambda(t,r)}
  f(t,r,w,L)\, dr\, dw\, dL,
\end{align} 
while in maximal areal coordinates,
\begin{align} \label{eq:def_Nma(t)}
  N = 4 \pi^2 \int_0^\infty \int_{-\infty}^\infty \int_0^\infty a(t,r)
  f(t,r,w,L)\, dr\, dw\, dL,
\end{align}
and in Eddington-Finkelstein coordinates,
\begin{align} \label{eq:def_Nef(t)}
  N = 4 \pi^2 \int_0^\infty \int_0^\infty \int_0^\infty
  f(t,r,p_1,L)\, dr\, dp_1\, dL .
\end{align}
The weights $e^\lambda$ and $a$ appear because the characteristic
flow is not measure preserving. This is due to the use of
non-canonical momentum variables. In fact, if by $D$ we denote
differentiation along a characteristic of the Vlasov equation,
then in the Schwarzschild case 
\[
D(f\, dx\, dv) = -\left ( \dot \lambda +
\lambda' e^{\mu-\lambda} \frac w \varepsilon \right )\, f\, dx\, dv
\]
and in the maximal areal case
\[
D(f\, dx\, dv) = -\left ( \frac{\alpha a'}{a^2} \frac w \varepsilon +
\beta' + \frac{2\beta}{r} \right)\, f\, dx\, dv.
\]
In Eddington-Finkelstein coordinates canonical momentum variables
are used which implies $D( f\, dx\, dv) = 0$. 

Another important difference between the three coordinate systems
is the existence of a criterion for the formation of trapped surfaces.
There exists no such criterion in Schwarzschild coordinates since
these coordinates cannot cover an open region which contains a trapped surface.
In maximal areal coordinates a trapped surface is present when
\begin{align}\label{eq:TS_cond_MA}
\frac 1 {a(t,r)} - r \kappa(t,r) < 0 .
\end{align}
In this case, the expansion of both outgoing and ingoing null geodesics
is negative at the time $t$ on the sphere of radius $r$. This signals
the development of a spacetime singularity, cf.\ \cite{Pen1965}.
In Eddington-Finkelstein coordinates the condition 
\begin{equation}\label{eq:TS_cond_EF}
a(t,r) < 0 
\end{equation}
corresponds to the existence of a trapped surface, cf.~\cite{AnRe2010}.

In Schwarzschild coordinates there exists a local existence and uniqueness
result for smooth, compactly supported initial data together with a
continuation criterion for such solutions, cf.\ \cite{Rein95,RR92a}.
An analogous result holds in maximal areal coordinates, cf.\ \cite{Gue19},
but probably not in Eddington-Finkelstein coordinates, at least not for
general smooth data whose support contains the origin. In general,
Schwarzschild or maximal areal coordinates are more useful when proving
that certain data launch global, geodesically complete solutions,
cf.\ \cite{AKR1,RR92a}, while the formation of trapped surfaces for
suitable data has been shown in Eddington-Finkelstein coordinates,
cf.\ \cite{AnRe2010}. The stability analysis in
\cite{HaRe2013,HaRe2014,HaLiRe2020} was carried out in Schwarzschild
coordinates.

\subsection{Steady state solutions} \label{ssc:steady}

Despite the differences of the above coordinate systems,
there exists an explicit coordinate transformation which maps
stationary solutions in one coordinate system into stationary
solutions in the other two. For this reason we discuss
steady states in Schwarzschild coordinates first.

A simple calculation shows that for a time-independent metric
the particle energy 
\begin{align*}
  E = \begin{cases}
    \varepsilon e^{\mu} & \text{for Schwarzschild,} \\
    \varepsilon  \alpha & \text{for maximal areal,} \\
    \frac{b}{2}\Big( a p_1 + \frac{1+L/r^2}{p_1} \Big) &
    \text{for Eddington-Finkelstein} 
  \end{cases}
\end{align*} 
is conserved along characteristics of the Vlasov equation,
and due to spherical symmetry the same is true for
the square of the angular momentum $L$.
Thus, every ansatz of the form $f_0 = \phi(E,L)$ solves the Vlasov
equation in the time-independent metric and reduces the
system to the field equations for that metric, where the source
terms now depend on the latter through the given ansatz.
In the present paper we consider two different types of ansatz functions,
namely the polytropic ansatz
\begin{align} \label{eq:Ansatz_Poly}
f_0 = \left( 1 - \frac{E}{E_0} \right)^k_+ \left( L - L_0 \right)^l_+
\end{align}
and the King model
\begin{align} \label{eq:Ansatz_King}
f_0 = \left( e^{ 1 - \frac{E}{E_0}} - 1 \right)_+.
\end{align}
Here $z_+$ denotes the positive part of some number $z$.
The constants $E_0 > 0$ and $L_0 \geq 0$ are the cut-off energy and
the minimal angular momentum while $k \geq 0 $ and $l \geq 0$ are
prescribed parameters. In order to guarantee the existence of
stationary solutions with finite ADM mass and compact support
\cite{RaRe2013}, we choose $k < l + \frac32$, but also consider
the borderline case of $k=\frac 3 2$ and $l=0$. The cut-off energy
is necessary to ensure finite extension and finite ADM mass of the
corresponding stationary solution. Prescribing $L_0 > 0$ gives rise
to a vacuum region at the center and hence to a shell type solution.
On the other hand, choosing $L_0 = 0$ implies that the support of the
steady state contains the origin.
Note that the factor $\left( L - L_0 \right)^l_+$ can also be multiplied
to the King ansatz, with the analogous effect on the support of the resulting
steady state. 

Even though multiple radially separated shells may arise from a
polytropic ansatz with $L_0>0$,
we only consider the innermost shell as our steady state,
cf.\ \cite{AnRe2007}. Formally, these steady states depend not only on $E$
and $L$ but also on the choice of the shell. However, we still refer to them
as polytropes. 

In order to solve for the metric coefficients, it turns out to
be more convenient to consider $y = \ln E_0 - \mu$. For an ansatz
as above we can write the spatial mass density and pressure as
functions of $y$, i.e., 
\[
\rho (r) = g(y(r)), \quad p(r) = h(y(r)).
\]
For the definition of $g$ and $h$ and further details we refer to
\cite{RaRe2013}. The stationary Einstein-Vlasov system is then reduced
to the differential equation
\begin{equation}\label{eq:ssdgl}
  y' = - \frac{1}{1-\frac{8\pi}{r}\int_0^r  g(y(s))s^2 \, ds}
  \left( \frac{4\pi}{r^2} \int_0^r g(y(s)) s^2 \, ds + 4\pi r h(y(r)) \right).
\end{equation}
This equation is \eqref{eq:mu'} with \eqref{eq:exp(-2lambda)} substituted in.
For every choice of \mbox{$y(0) = y_0 > 0$}, \cite{RaRe2013} guarantees a unique
solution of \eqref{eq:ssdgl} with finite ADM mass and compact support.
Hence one given ansatz of the
form \eqref{eq:Ansatz_Poly} or \eqref{eq:Ansatz_King}
yields a one-parameter family of such steady states.
The parameter is
closely connected to the central redshift $z_c$, measuring the redshift
of a photon which is emitted at the center $r = 0$ and received at the
boundary of the steady state:
\begin{align} \label{eq:def_central redshift}
z_c =  e^{y_0} - 1.
\end{align}
This is not the standard definition of the central
redshift where the photon is received at infinity,
but our definition is more suitable here.
If $y$ is a solution to \eqref{eq:ssdgl}
and the cut-off energy is defined as $E_0=\lim_{r\to\infty}e^{y(r)}$,
the metric coefficient $\mu = \ln E_0 - y $ satisfies the proper boundary
condition at infinity, and $\lambda$ is given by
\[
\lambda(r)  = - \frac 1 2  \,
\mathrm{ln} \left ( 1-\frac{8\pi}{r} \int_0^r \rho(s) s^2 ds \right ) .
\]
We now discuss how to obtain the stationary solutions in the other
two coordinate systems. In maximal areal coordinates stationary
solutions satisfy $\kappa =\beta= j = 0$. The metric is then equivalent
to the Schwarzschild metric by simply setting $\mu = \mathrm{ln} \, \alpha$
and $\lambda = \mathrm{ln} \, a$, cf.\ \cite{AnRe2006}.
In the case of Eddington-Finkelstein coordinates the metric coefficients
$a$ and $b$ can be obtained by the change of variables
\begin{align*}  
t \mapsto t + \int_0 ^r e^{\lambda(s) - \mu(s)} \, ds,
\end{align*}
and the metric coefficients are related via
\begin{align*}
a= e^{-2 \lambda}, \quad b= e^{\lambda + \mu  }.
\end{align*}  
Before we discuss how we solve the time-dependent Einstein-Vlasov system, we briefly
explain how we compute the steady states.
For a prescribed ansatz function of the form \eqref{eq:Ansatz_Poly}
or \eqref{eq:Ansatz_King} as well as $y_0 > 0$, we compute the corresponding
solution $y$ to the equation \eqref{eq:ssdgl}.
We use the explicit Euler method enhanced with a leap frog scheme
to solve (\ref{eq:ssdgl}) and apply Simpson's rule to calculate the
integrals appearing in $g$, $h$, and (\ref{eq:ssdgl}).
Note that the whole steady state computation has to be
done only once, which allows us to use very high accuracy for
this part. The distribution function $f_0$ of the
steady state is then given via \eqref{eq:Ansatz_Poly}
or \eqref{eq:Ansatz_King}.

\section{The numerical method for the time evolution} \label{sc:numeric} 
\setcounter{equation}{0}

The algorithms used to investigate the stability of steady states of
the above system are based on the particle-in-cell scheme.
This scheme has also been used in \cite{AnRe2006,RaRe2018,ReReSch}.
For the spherically symmetric Vlasov-Poisson system its convergence
has been shown in \cite{Sch}, the analogous result for the Einstein-Vlasov
system in Schwarzschild coordinates is shown in \cite{ReRo}.
In a particle-in-cell scheme the support of the distribution
function $f$ is initially split into distinct cells.
Into each cell a numerical
particle is placed to represent the contribution to $f$ of this cell,
and these particles are then propagated according to the
Einstein-Vlasov system. Most of the following steps differ for the
three coordinate systems under investigation. We focus on the algorithm
for Schwarzschild coordinates, but always highlight difficulties
arising in the other two coordinate systems and how to overcome them.

To initialize the numerical particles we use variables adapted
to the spherical symmetry of the steady state $f_0$, i.e., we write
\begin{align*}
f_0 = f_0(r,w,L)=f_0(r,u,\psi),
\end{align*}
where we use the additional variables $u\geq0$ and $\psi\in[0,\pi]$ given by 
\begin{align*}
u^2 = w^2 + \frac L{r^2} ,\quad w = u\cos(\psi).
\end{align*}
Assuming that $f_0$ vanishes outside of the set
$[R_-,R_+]\times[U_-,U_+]\times[\Psi_-,\Psi_+]$,
we prescribe a radial step length $\Delta r > 0$
as well as integers $N_u$ and $N_\psi$ to define the step lengths
\begin{align*}
\Delta u = \frac{U_+-U_-}{N_u} ,\quad \Delta\psi=\frac{\Psi_+-\Psi_-}{N_\psi}
\end{align*}
and set up a grid of points 
\begin{align*}
  r_i = \left(i-\frac12\right)\Delta r,\quad u_j
  =\left(j-\frac12\right)\Delta u,\quad\psi_k=\left(k-\frac12\right)\Delta\psi.
\end{align*}
At each point $(r_i,u_j,\psi_k)$ we generate a numerical particle
carrying the weight
\begin{align*}
  f_{i,j,k} = f_0(r_i,u_j,\psi_k) \,
  4\pi r_i^2\Delta r \, 2\pi u_j^2\Delta u \, \sin(\psi_k)\Delta\psi,
\end{align*}
where $f_0(r_i,u_j,\psi_k)$ is calculated using the steady state
from above. We use $(r,u,\psi)$-variables in Schwarzschild and
maximal areal coordinates to generate the numerical particles.
In the Eddington-Finkelstein case a similar initialization scheme
based on $(r,p_1,L)$-variables is used. For the following steps
it is convenient for Schwarzschild and maximal areal coordinates
to write the particle positions in $(r,w,L)$-variables, since $L$
is conserved along characteristics. 

We then compute the matter quantities $\rho$, $p$, and $j$ on the
fixed radial grid given by $r_j=j\Delta r$ by integrating $f_0$
according to \eqref{eq:rho(r,w,L)}, \eqref{eq:p(r,w,L)}, and
\eqref{eq:j(r,w,L)}. This is implemented by adding up $f_{i,j,k}$
with the appropriate weight given by
\eqref{eq:rho(r,w,L)}-\eqref{eq:j(r,w,L)} and by a linear interpolation
of $f_0$ in the radial direction; note that $f_{i,j,k}$ contains the phase
space volume element. 

Next, we compute the Hawking mass $m$ on the fixed radial grid by using
\eqref{eq:def_m(r)_Hawking mass} and a quadratic interpolation, taking
into account the possible order of the integrand near the origin.
Afterwards, the metric quantity $\mu$ is calculated by applying
\eqref{eq:exp(-2lambda)}, \eqref{eq:mu'}, and the boundary condition
$\mu(R)=\lambda(R)$, where $R$ denotes the outer boundary of the radial
support. 

In maximal areal coordinates we instead solve for $\eta$ and $\kappa$
simultaneously by employing a fourth-order Runge-Kutta method in order
to guarantee sufficient numerical precision and to appropriately handle
the implicit structure of the field equations. Determining
$\alpha$ requires more effort due to the ellipticity of equation
\eqref{eq:iv} and the boundary condition given at spatial infinity.
We choose a sufficiently large grid, approximate the various derivatives
and arrive at a tridiagonal system which can be solved explicitly.
For more details consider the scheme used in \cite{AnRe2006}.

The time step is then performed by propagating the numerical particles
according to the characteristic system corresponding
to the Vlasov equation. In order to avoid numerical errors at the spatial
origin caused particularly by particles coming close to the origin, it is
advantageous to use Cartesian coordinates for the propagation. Here, all
functions involved are interpolated according to their order, especially
near the origin. However, we do not use Cartesian coordinates in the case
of Eddington-Finkelstein coordinates since the metric written in these
coordinates is not continuous at the origin, cf.\ \cite{AnRe2010}. The new
position of each particle is then computed by a proper stepping
method with a prescribed time step size $\Delta t > 0$. Note that we also
have to update the weight $f_{i,j,k}$ during each time step in the case of
Schwarzschild and maximal areal coordinates, since the characteristic flow
of the Vlasov equation
system is not measure preserving in these coordinates. 

At this point the particle coordinates and weights are known at the
time $\Delta t$ and we can repeat the iteration until we reach
some final time $T$ which is prescribed from the beginning.

So far we have numerically evolved a steady state $f_0$ itself,
but in order to analyze its stability properties
we have to perturb it. In \cite{AnRe2006} this is done by taking $A\,f_0$
for some parameter $A\approx 1$ as an initial condition.
However, this kind of perturbation is not natural from a physics point
of view since it is not dynamically accessible, i.e., it does not
preserve Casimir functionals of the form
\begin{align}
\mathcal C (f(t)) = \iint e^{\lambda(t,x)} \chi( f(t,x,v)) \,dx\,dv \label{eq:casimir}
\end{align}
where $\chi\in C^1(\R)$ with $\chi(0)=0$; a physically viable perturbation
for example by some external force should preserve these.
We provide a perturbation procedure preserving these invariants.
An analogous type of perturbation has been used in
\cite{RaRe2018} for the Vlasov-Poisson system, but as far as we know
this is the first implementation of dynamically accessible
perturbations to steady states of the Einstein-Vlasov system.
During an initial time interval $[0,T_{pert}]$ we propagate the
numerical particles according to the modified characteristic system
\begin{align}\label{eq:perturb}
  \dot x =  e^{\mu - \lambda} \frac v\varepsilon ,\quad \dot v
  = - \left( \dot\lambda \frac{x\cdot v}r +
  \mu' e^{\mu - \lambda} \varepsilon \right) \frac xr + \gamma \frac xr 
\end{align}
in the Schwarzschild case for prescribed $\gamma\approx0$.
Compared to the original characteristic system, we add the
term $\gamma \frac xr$ to the right hand side of the equation
for $\dot v$. Since Casimir functionals are preserved along
solutions of the Vlasov equation and the above perturbation
does not contribute to the divergence of the right hand side
of the characteristic system, this perturbation is indeed
dynamically accessible. In maximal areal coordinates we also
add the divergence free term $(0,\gamma \frac xr )$ to the right
hand side of the respective characteristic system during an
initial time interval. In Eddington-Finkelstein coordinates,
where $E=E(r,p_1)$ is the Hamiltonian governing the motion of
the particles, we add $\gamma rp_1$ as a perturbation of the
Hamiltonian.

Despite the fact that $v$ does not denote the canonical momentum,
all these perturbations can be interpreted as particles being
accelerated either radially outwards if $\gamma >0$ or radially
inwards if $\gamma < 0$. 

We refer to the evolved state at $t=T_{pert}$ as the perturbed state.
To determine the intensity of the perturbation we consider
$e^{\mu(t,0)}$,  $\alpha(t,0)$, and $b(t,0)$ respectively. We prescribe
a small number $\epsilon_{pert}$
and then choose $\gamma$ such that the relative error of
$e^{\mu(t,0)}$ between $t=0$ and $t=T_{pert}$ is close to $\epsilon_{pert}$.
Obviously, numerical and evolutionary effects contribute to the
perturbation of the steady state as well. To this end, we choose
$\epsilon_{pert}$ such that it dominates the relative error of the
above quantities resulting from the numerical initialization. In addition,
we choose $T_{pert}$ sufficiently small such that evolutionary effects
are negligible. 

We emphasize that we have tested several other perturbations.
On the one hand, we used rather simple perturbations where we
scale the steady state by some factor or shift it slightly in
one direction. On the other hand, we employed further dynamically
accessible perturbations where we propagate the numerical particles
using modified metric quantities during an initial time interval.
For every kind of perturbation there seem to exist precisely two distinct
characteristic behaviors: pushing the steady state either towards
collapse or towards dispersion. However, all the effects described
in the next section seem to only depend on the direction of
the perturbation but not on its specific type. 

Let us end this section with a remark concerning the numerical implementation.
In order to obtain reliable physical results, we have to work with tens
of millions of numerical particles and sufficiently small time steps. For these simulations to run within
a reasonable time-frame, the programs have to be parallelized,
which however fits very well with the particle-in-cell scheme.
We refer to \cite{KoRaRe2013} for a detailed discussion.
\section{Results}\label{sc:results}
\setcounter{equation}{0}
In the following we investigate the stability behavior of the
one-parameter families of steady states which we obtain by choosing
the King model or the polytropic ansatz with fixed parameters
$k$, $l$, and $L_0$. Due to its physical relevance we mainly present
the results for the King model. The parameter that determines the
steady state is the central value $y_0$ where larger values of $y_0$
generally indicate more relativistic scenarios, cf.\
\eqref{eq:def_central redshift}. For a detailed overview of
properties of the steady states we refer to \cite{AnRe2007}. 

\begin{figure}[h]
	\begin{center}
		\input{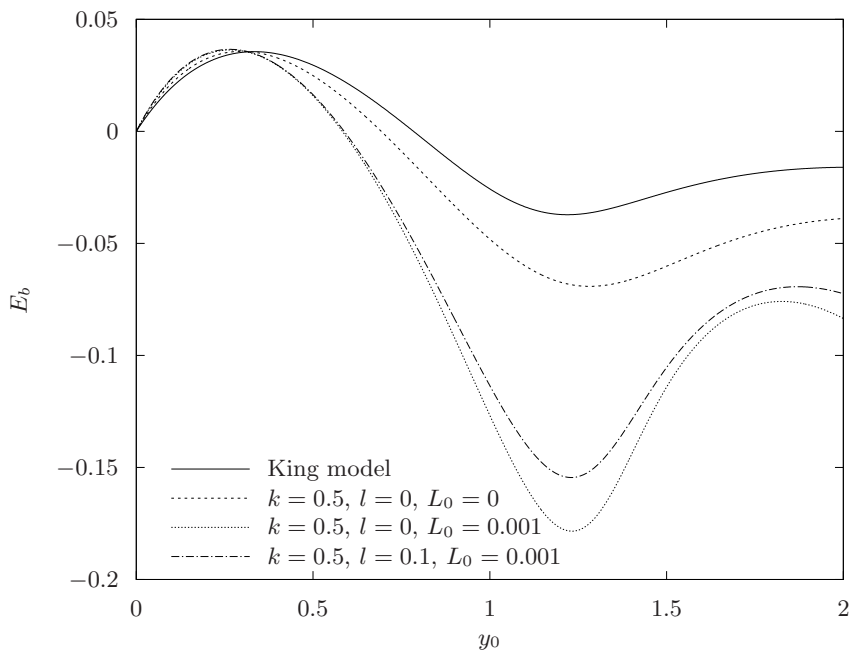}
	\end{center}
	\vspace*{-.7cm}\caption{Binding energy for different models.}
	\label{img:bindeng}
\end{figure}

\begin{figure}[H]
	\begin{center}
		\input{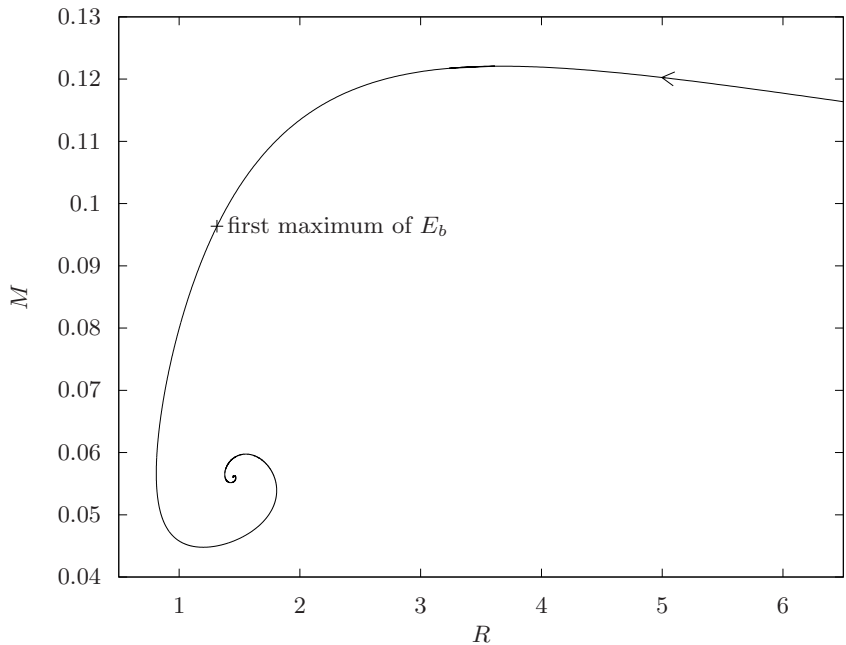}
	\end{center}
	\vspace*{-.7cm}\caption{Mass-radius spiral for the King model; as $y_0$ increases the corresponding $(R,M)$ moves into the spiral.}
	\label{img:spirale}
\end{figure}

A physically meaningful quantity of a steady state is its binding energy
\[
E_b = \frac{N - M}{N},
\]
where $N$ is the number of particles and $M$ the ADM mass.
In Figure~\ref{img:bindeng} the binding energy is plotted against
$y_0$ for different models, namely for the King model and polytropes
with $(k,l,L_0) = (0.5, 0, 0)$, $(0.5, 0, 0.001)$, $(0.5, 0.1, 0.001)$.
We observe that the binding energy has the same qualitative
features across the different models: It develops
a positive local maximum after which it drops below zero.
Despite the fact that the parameter \mbox{$L_0=0.001$} may seem very small,
it still suffices to change the corresponding one-parameter
family of steady states significantly. Table~\ref{tab:bindengmax} gives the values
$y_{\mathrm{max}}$ and $y_{\mathrm{zero}}$
where the binding energy attains its first maximum or
its first zero respectively for the different models used
in Figure~\ref{img:bindeng}.
\begin{table}[h]
	\begin{center}
		\begin{tabular}{| l | c | c |}
			\hline
			Model & $y_{\mathrm{max}}$ & $y_{\mathrm{zero}}$ \\\hline
			King & 0.334 & 0.784 \\
			$k=0.5$, $l=0$, $L_0=0$ & 0.298 & 0.693 \\
			$k=0.5$, $l=0$, $L_0=0.001$  & 0.267 & 0.583 \\
			$k=0.5$, $l=0.1$, $L_0=0.001$  & 0.265 & 0.588 \\\hline
		\end{tabular}
		\caption{First binding energy maximizer and zero. }\label{tab:bindengmax}
	\end{center}
\end{table} %TODO Diskussion: Genauigkeit angeben?
Another way to visualize a one-parameter family of steady states
originating from a given ansatz function is the so-called mass-radius
diagram where for each value of $y_0>0$ one plots the ADM mass and radius
of the corresponding steady state. In Figure~\ref{img:spirale}
this is done for the King model.

The spiral structure is a general feature of these mass-radius diagrams
for the Einstein-Vlasov system, cf.\ \cite{AnRe2007}. According to the
so-called turning point principle the steady state should pass from
being stable to being unstable as it crosses the first maximum point
along the curve, but we find that this is not true.
As mentioned at the end of Section~\ref{sc:numeric}, perturbations
can be categorized as either promoting collapse or promoting dispersion.
For example, the dynamically accessible perturbation of $f_0$ with
parameter $\gamma$ described in \eqref{eq:perturb} pushes the steady
state towards collapse if $\gamma<0$, and towards dispersion if $\gamma>0$.

\subsection{Stable steady states}\label{ss:stabless}
\begin{figure}[h]
  \begin{center}
    \input{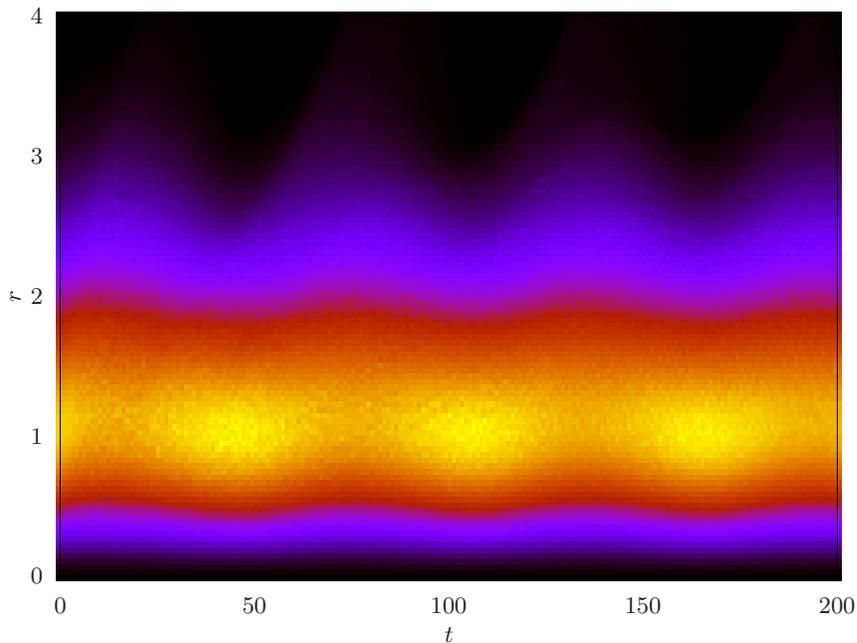}
  \end{center}
  \vspace*{-.7cm}
  \caption{The oscillation of the weighted mass density $4\pi r^2 \rho$ for the King model in
    Schwarzschild coordinates for $y_0=0.1$
    perturbed by a dynamically accessible perturbation with direction
    $\gamma>0$. The color
    represents the value of $4\pi r^2 \rho$, increasing from black to yellow.}
  \label{img:rho_osc}
\end{figure}

\begin{figure}[h]
  \begin{center}
    \input{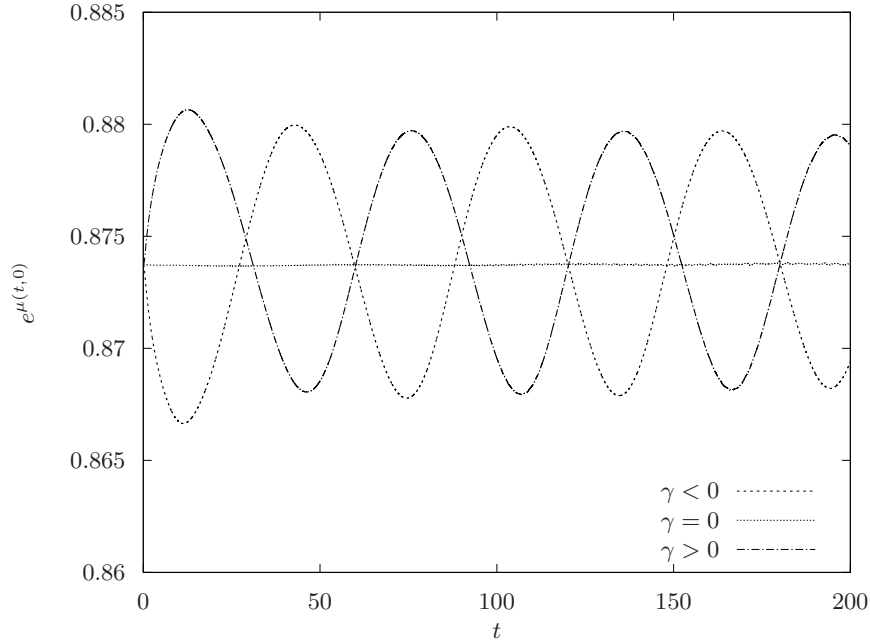}
  \end{center}
  \vspace*{-.7cm}
  \caption{The King model in Schwarzschild coordinates for $y_0=0.1$
    perturbed by dynamically accessible perturbations with directions
    $\gamma<0$, $\gamma=0$, and $\gamma>0$. }
  \label{img:kingconstandosc}
\end{figure}

For small values of $y_0$ we find that the steady states are stable
with respect to every reasonable perturbation. However, perturbations
do not leave the steady state unchanged but make it oscillate in a
pulsating manner. Similar oscillations have been numerically observed
as perturbations of stable steady states of the
Vlasov-Poisson system, cf.\ \cite{RaRe2018}. 
Figure~\ref{img:rho_osc}
illustrates this behavior on the level of the mass distribution.
Alternatively, the quantity $e^{\mu(t,0)}$, $\alpha(t,0)$, or $b(t,0)$, respectively,
is a good indicator of an oscillation---or other solution behaviors---since it is the relativistic
counterpart of the gravitational potential at the spatial origin.
In addition, these quantities are integrated from infinity inwards
which makes them less volatile from a numerical point of view
while they are influenced by the solution as a whole. 
For a comparison of the effect of different signs of $\gamma$
when perturbing a stable steady state consider
Figure~\ref{img:kingconstandosc}.
On the one hand, for a perturbation that promotes collapse
($\gamma < 0$) the quantity $e^{\mu(t,0)}$ initially decreases.
On the other hand, $e^{\mu(t,0)}$ initially increases for a
perturbation that promotes dispersion ($\gamma > 0$).
In both cases an oscillation of the perturbed steady state develops.
Concerning the period of such oscillations, we observe that it depends
on the specific steady state model as well as the strength of the
perturbation itself. The period seems to approach a
fixed positive value as the strength of the perturbation is gradually
decreased, but in the context of stability issues it is not relevant
to determine the periods of perturbed stable steady states. 

Across all models and coordinate systems, we observe that the
steady states are stable for small values of $y_0$.
At some threshold value the stability behavior changes.
\subsection{The first binding energy maximum---onset of instability}
\label{sc:bindingenmaxhypo}
One main goal of our numerical investigation of the time dependent system
is to detect when instability of steady states first occurs along
a one-parameter family.
We find convincing evidence that the so-called \enquote{binding energy maximum
hypothesis} holds which has also been analyzed and confirmed
in \cite{AnRe2006,Ip1969,Ip1980,ShTe1985} for different models
and perturbations. It was first proposed by Zel'dovich et al.\ in
\cite{Ze1966, Ze1971}. The hypothesis states that the first
binding energy maximum along a steady state sequence signals
the onset of instability.

The starting point of our investigation is \cite{AnRe2006}
where the binding energy maximum hypothesis has been verified
numerically in maximal areal coordinates for the case of polytropic
shell steady states, i.e., $L_0 > 0$ in \eqref{eq:Ansatz_Poly}. 
We are able to confirm the findings of \cite{AnRe2006}
in all three coordinate systems. Furthermore, we expand these results
to polytropes with no inner vacuum region, i.e., $L_0=0$, and in
particular to isotropic models, i.e., $L_0 = 0$ and $l=0$,
which allows particles to pass through the origin.
Our findings clearly support the binding energy maximum hypothesis
in these cases.

\begin{figure}[h]
  \begin{center}
    \input{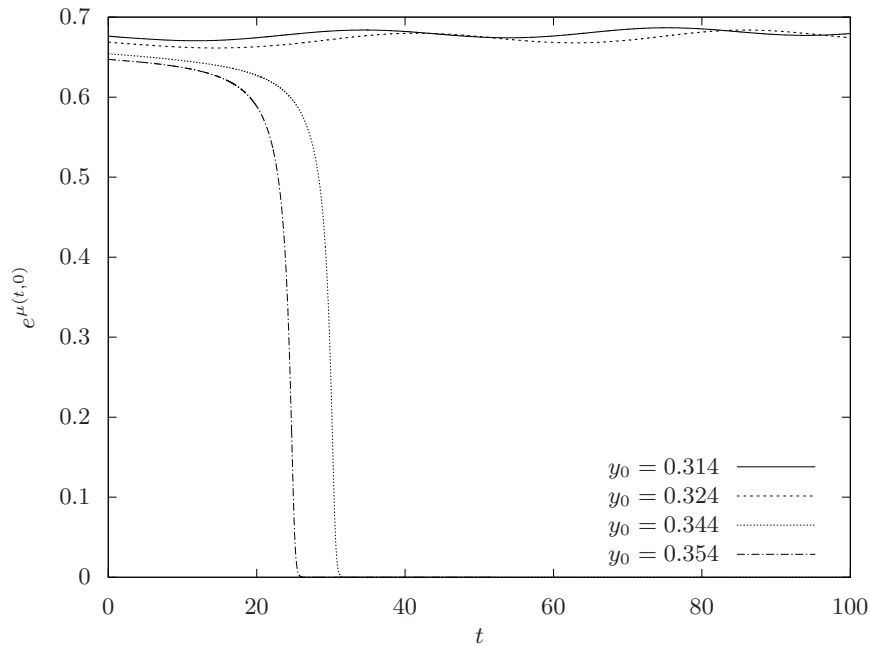}
  \end{center}
  \vspace*{-.7cm}\caption{The King model in Schwarzschild coordinates
    for $y_0$ close to $y_{\mathrm{max}}$ perturbed by a dynamically
    accessible perturbation with direction $\gamma<0$. }
  \label{img:kingmaxcol}
\end{figure}

We present the results of simulations for
four cases across the three coordinates
systems: the isotropic case for $k=0.5, \, 1, \, 1.5$ and the King model.
In every case we determine the first maximizer of the binding energy,
defined as $y_{\mathrm{max}}$, with sufficient accuracy. We then consider
steady states with values of $y_0$ close to $y_{\mathrm{max}}$.
As mentioned in Section~\ref{ss:stabless} we find that the steady states
are stable for both types of perturbations for $y_0 < y_{\mathrm{max}}$
across all models and coordinate systems.
For $y_0 > y_{\mathrm{max}}$ the stability behavior changes.
In fact, in all coordinate systems and models we observe that a
perturbation promoting collapse leads to the actual collapse of
the steady state to a black hole,
which is illustrated in Figure~\ref{img:kingmaxcol}.
Notice that for $y_0 < y_{\mathrm{max}}$ we observe periodic oscillations
even when $y_0$ is very close to $y_{\mathrm{max}}$.
We elaborate on the details of a collapse in the next section.

\begin{figure}[h]
  %\centering
  \begin{subfigure}[h]{0.49\textwidth}
    \centering
    \resizebox{!}{.72\textwidth}{\input{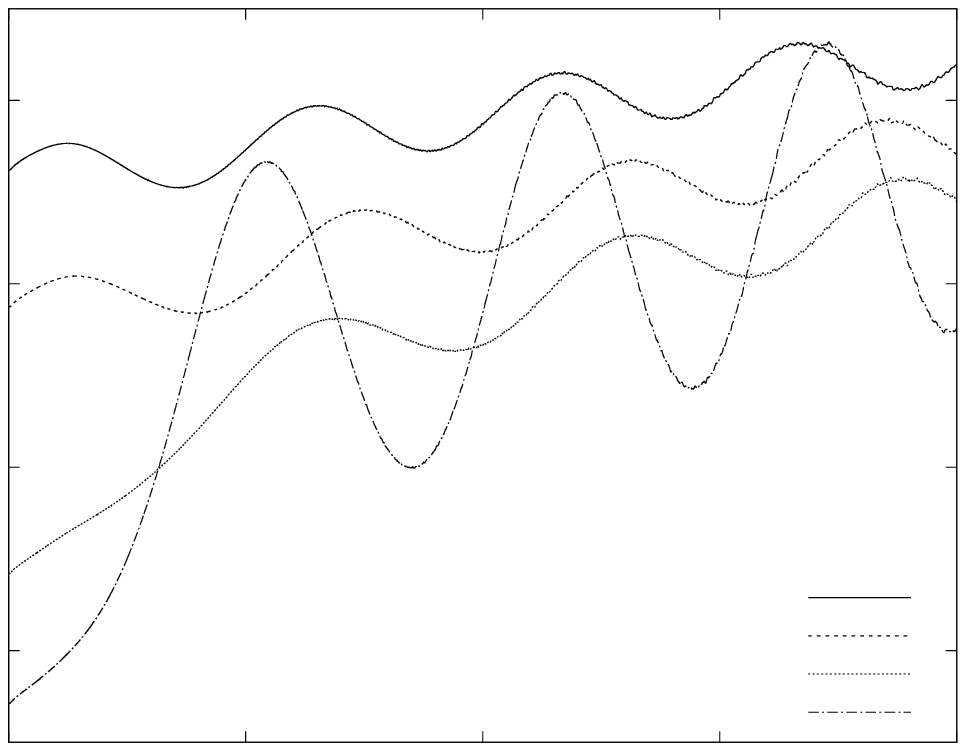}}\vspace*{-.16cm}
  \end{subfigure}
  \begin{subfigure}[h]{0.49\textwidth}
    \centering
    \resizebox{!}{.72\textwidth}{\input{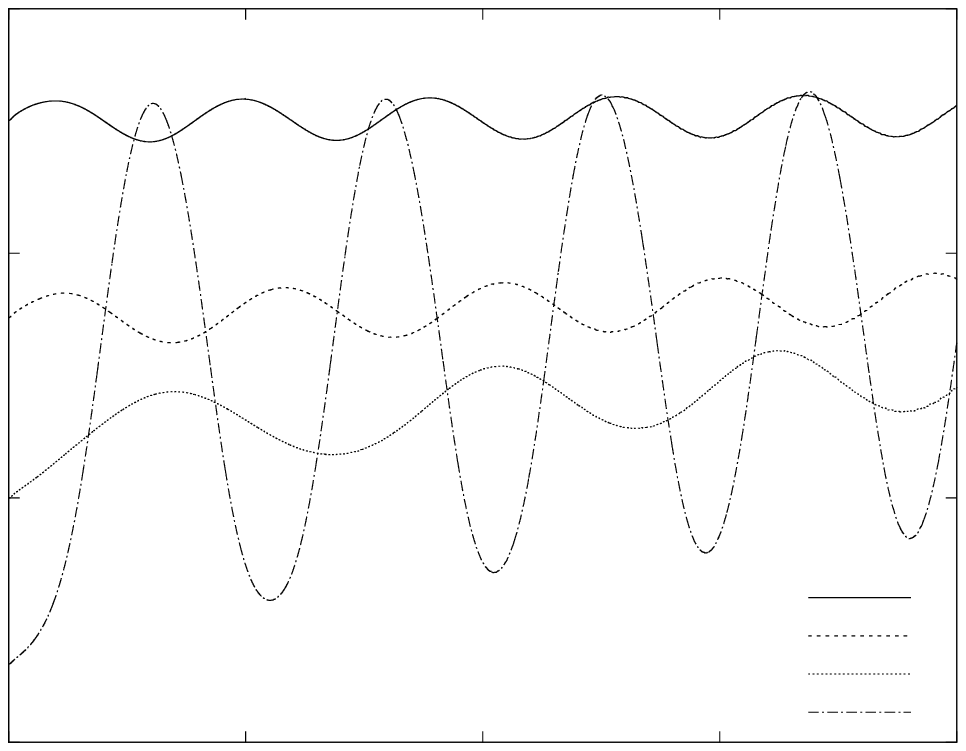}}\vspace*{-.16cm}
  \end{subfigure}
  %\vspace*{-1cm}
  \caption{The King model in Schwarzschild coordinates for
    $y_0$ close to $y_{\mathrm{max}}$ perturbed by a dynamically
    accessible perturbation with direction $\gamma>0$. Notice the different scales.}
  \label{img:kingmaxdis}
\end{figure}

Perturbations promoting dispersion neither lead to collapse nor
full dispersion but invoke an oscillation of the system.
This may lead one to believe that the steady states for $y_0 > y_{\mathrm{max}}$
are in fact stable with respect to the dispersive perturbation.
However, by taking a closer look at $e^{\mu(t,0)}$ in these cases,
we notice that the oscillating quantity seems to drift upwards
which contradicts the perception of stability, i.e.,
an oscillation around the original steady state.
This behavior is shown in Figure~\ref{img:kingmaxdis}.
The attentive reader will have noticed that
in Figure~\ref{img:kingmaxdis} the aforementioned drift
seems to appear for $y_0 < y_{\mathrm{max}}$ close to $y_{\mathrm{max}}$
as well. However, further simulations showed that the latter
drift is an artifact of mere numerical inaccuracy.
For example, the drift weakens when decreasing the time increment $\Delta t$.
This might be an effect related to the well-known energy drift for
Hamiltonian systems. 

The crucial observation is that for \mbox{$y_0 > y_{\mathrm{max}}$}
the initial elevation of $e^{\mu(t,0)}$ significantly differs from
the initial behavior for \mbox{$y_0 < y_{\mathrm{max}}$}. On the one hand,
for $y_0=0.314$ and $y_0=0.324$ in Figure~\ref{img:kingmaxdis}, i.e., $y<y_{\mathrm{max}}$, the oscillation starts immediately and
the first local maximum is at about $t=15$. On the other hand, for
$y_0=0.344$ and $y_0=0.354$ there exists an initial phase over the
course of which the increase of $e^{\mu(t,0)}$ seems to occur
independently from the oscillation. Therefore, the first local
maximum appears much later at about $t=50$. The general behavior for
$y_0 > y_{\mathrm{max}}$ will be analyzed more
closely in Section~\ref{res:orbits}.

We now come back to the question whether the turning point principle
is valid for the Einstein-Vlasov system. As explained above, our numerics show
that in a one-parameter family steady states are stable as long as
$y_0 < y_{max}$, and they become
unstable as soon as $y_0 > y_{max}$; the change from stability to instability
occurs at the first maximum of the binding energy. In the mass-radius spiral
in Figure~\ref{img:spirale} we have marked the point which corresponds
to $y_{max}$, i.e., to the first maximum of the binding energy.
But this point is clearly to the left of the first turning point of
the mass-radius spiral, i.e., steady states for the Einstein-Vlasov system
remain stable even after the turning point principle would predict their
instability and after the very same steady states have become unstable
as steady states of the Einstein-Euler system. We have observed
this discrepancy or failure of the turning point principle in all cases
which we investigated numerically, and it seems a very interesting mathematical
problem to properly understand this issue. 
\subsection{Properties of the collapse}
One of the main advantages of using three coordinate systems is the
possibility to observe the properties of collapsing matter from different
perspectives. In Schwarzschild coordinates we have no analytical
criterion for the detection of a trapped surface. However, the
ratio $\frac{2m(t,r)}r$ approaching $1$ at some radius signals
the development of a black hole. On the contrary, in maximal areal
and Eddington-Finkelstein coordinates trapped surfaces can be
detected analytically, cf.\ \eqref{eq:TS_cond_MA} and \eqref{eq:TS_cond_EF}.
Note that rapidly increasing metric coefficients eventually cause
the program to break down in the event of a collapse.

\begin{figure}[h]
  \begin{center}
    \input{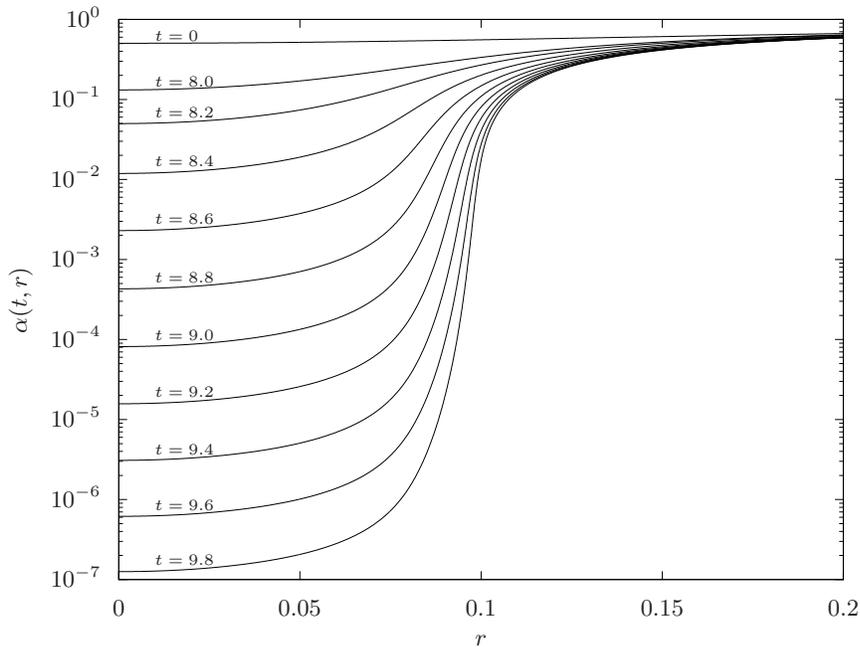}
  \end{center}
  \vspace*{-.7cm}
  \caption{Evolution of $\alpha(t,r)$ for the
    King model in maximal areal coordinates for $y_0=0.6$
    perturbed by a dynamically accessible perturbation with
    direction $\gamma<0$. Notice the logarithmic scale.}
  \label{img:collapselapse}
\end{figure}

\begin{figure}[h]
  \begin{center}
    \input{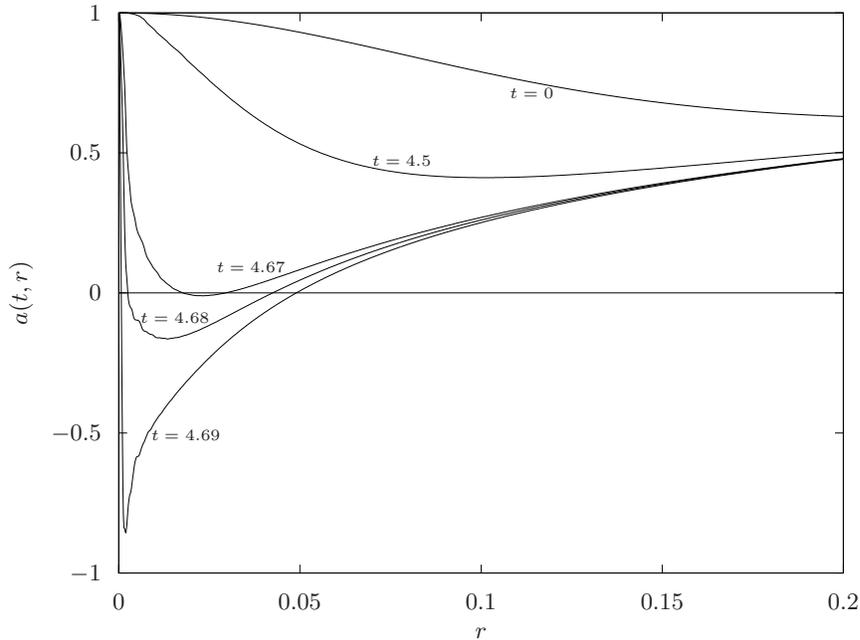}
  \end{center}
  \vspace*{-.7cm}
  \caption{Evolution of $a(t,r)$ for the King model in Eddington-Finkelstein
    coordinates for $y_0=0.6$ perturbed by a dynamically accessible
    perturbation with direction $\gamma<0$.}
  \label{img:acollapse}
\end{figure}

First, let us explain our observations when a collapse to a black hole
occurs. As far as the numerical particles within the simulation are
concerned, we notice that the matter focuses towards the origin and
the matter density $\rho$ increases at the center. In fact, in our
simulations no particles remain unaffected by the collapse, and all
particles are
eventually and irreversibly sucked towards the interior of the
trapped surface. This eventually leads to the radius $R$ of the
support of the solution tending to the Schwarzschild radius $2 M$.
This general behavior should be compared to the analytical findings
in \cite{AR}.
The observations on the particle level
can be linked to the characteristic behavior of
metric coefficients. In Schwarzschild and maximal areal coordinates
the collapse is reflected in the behavior of the functions
$e^{\mu}$ and $\alpha$, respectively. These quantities rapidly
decrease to zero in close proximity of the origin while still
satisfying the boundary conditions at infinity, cf.
\eqref{eq:boundary_conditions} and \eqref{eq:boundaryma}.
In the literature, this phenomenon is commonly known as
the \enquote{collapse of the lapse} since $e^{\mu}$ and $\alpha$
determine the amount of proper time which elapses from one
hypersurface of constant $t$ to the next. The collapse of the
lapse is depicted in Figure~\ref{img:collapselapse} where we
can see an exponential decay of $\alpha$ for late times and $r$
close to zero. In Eddington-Finkelstein coordinates, the metric
coefficient $b$ shows a similar behavior. 
Furthermore, the criteria for the formation of a trapped surface
mentioned above are satisfied from some moment in time onwards.
Figure~\ref{img:acollapse} shows the evolution of the metric
coefficient $a$ in Eddington-Finkelstein coordinates where $a<0$
corresponds to the occurrence of a trapped surface, cf.\
Section~\ref{ssc:comparison}.  In particular, the radius at which
the first trapped surface forms is significantly larger than the
chosen radial increment $\Delta r$. This supports the conjecture
that the weak censorship hypothesis holds for the Einstein-Vlasov
system in our setting. For an overview of this topic we refer
to \cite{An2011}.  
Independently of the choice of the coordinate system,
we observe that the coordinate time it takes until an unstable
steady state collapses decreases as $y_0$ is enlarged. This can
be explained by the fact that larger values of $y_0$ correspond to
more relativistic steady states. However, one should
recall that the time coordinate
$t$ has a different meaning across the three coordinate systems.
\subsection{Heteroclinic orbits} \label{res:orbits}
\begin{figure}[h]
  \begin{center}
    \input{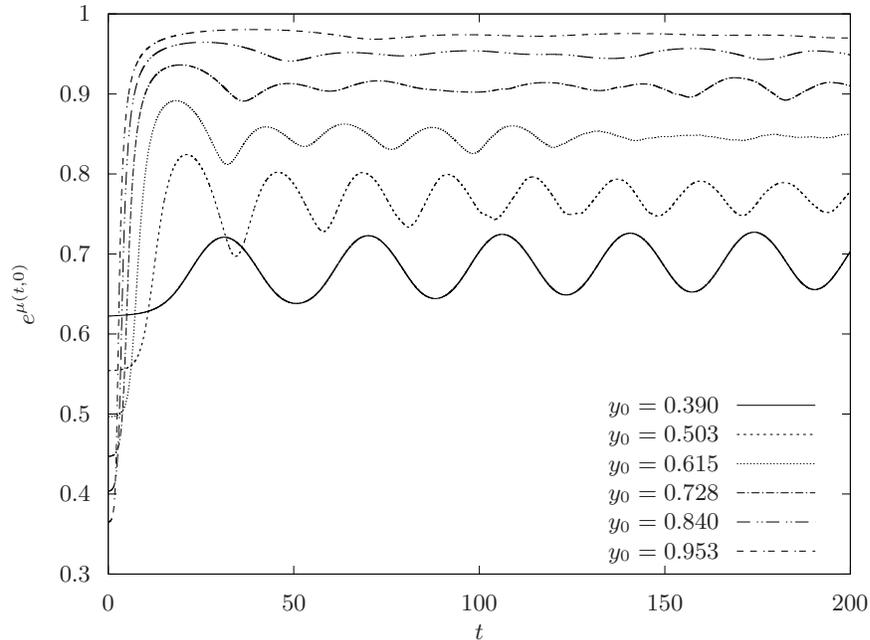}
  \end{center}
  \vspace*{-.7cm}
  \caption{The King model in Schwarzschild coordinates for
    $y_0 > y_{\mathrm{max}}$ perturbed by a dynamically accessible
    perturbation with direction $\gamma>0$.} 
  \label{img:kingdis}
\end{figure}

When perturbing an unstable steady state with a perturbation promoting
dispersion, i.e., $y_0>y_{\mathrm{max}}$ and $\gamma>0$, the resulting solution
disperses at first, which means that the matter distributes more evenly in
space and spacetime becomes flatter.
This corresponds to an initial increase of $e^{\mu(t,0)}$, $\alpha(t,0)$,
or $b(t,0)$, respectively. At least if $y_0$ is not too large, we can
clearly see that the solution starts reimploding at some point in time,
which means that $e^{\mu(t,0)}$ decreases again, giving rise to an
oscillating behavior. However, this oscillation occurs at values of
$e^{\mu(t,0)}$ larger than $e^{\mu(0,0)}$. We call this difference the
initial elevation. As indicated in Section~\ref{sc:bindingenmaxhypo},
the initial elevation can be observed for all $y_0 > y_{\mathrm{max}}$. 
These effects occur across all classes of steady states, types of
perturbations promoting dispersion, and coordinate systems in a
similar way and are illustrated in Figure~\ref{img:kingdis}
in the case of the King model in Schwarzschild coordinates for
various $y_0>y_{\mathrm{max}}$.
%\marginpar{\begin{color}{red} nur bis t=200!\end{color}}

\begin{figure}[h]
  \begin{center}
    \input{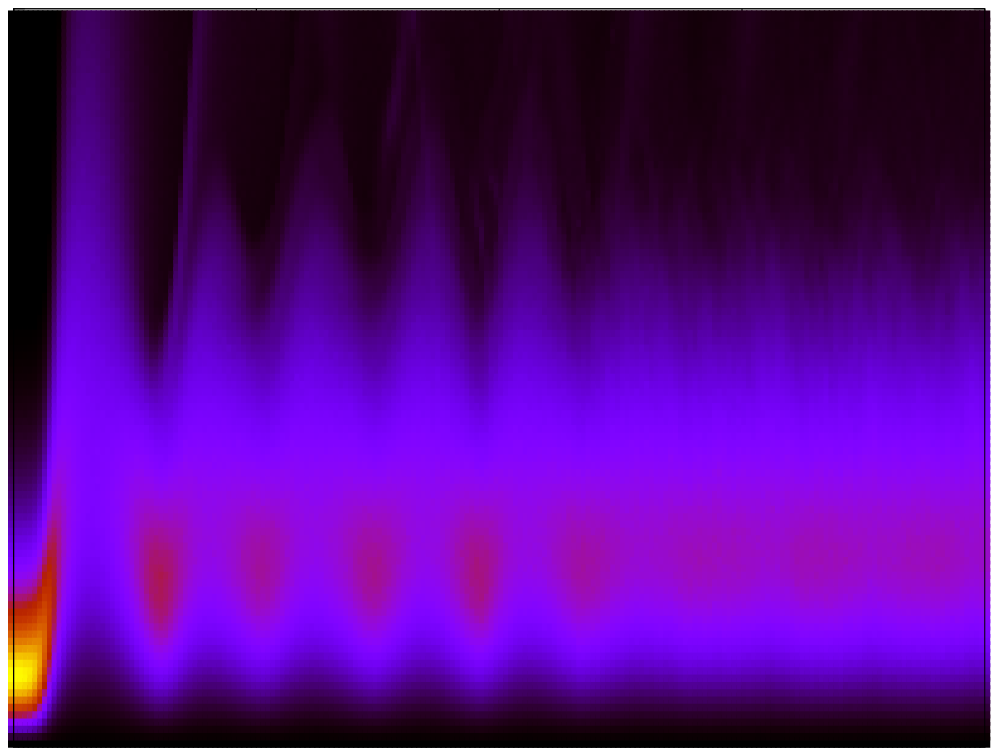}
  \end{center}
  \vspace*{-.7cm}
  \caption{The weighted mass density $4\pi r^2 \rho$ for the King model in Schwarzschild
    coordinates for $y_0 =0.6> y_{\mathrm{max}}$ perturbed by a dynamically
    accessible perturbation with direction $\gamma>0$. The color
    represents the value of $4\pi r^2 \rho$, increasing from black to yellow.} 
  \label{img:rhofalsemap}
\end{figure}

\begin{figure}[hp]
  \centering
  
  \begin{subfigure}[h]{0.49\textwidth}
    %\centering
    \resizebox{!}{.8\textwidth}{\input{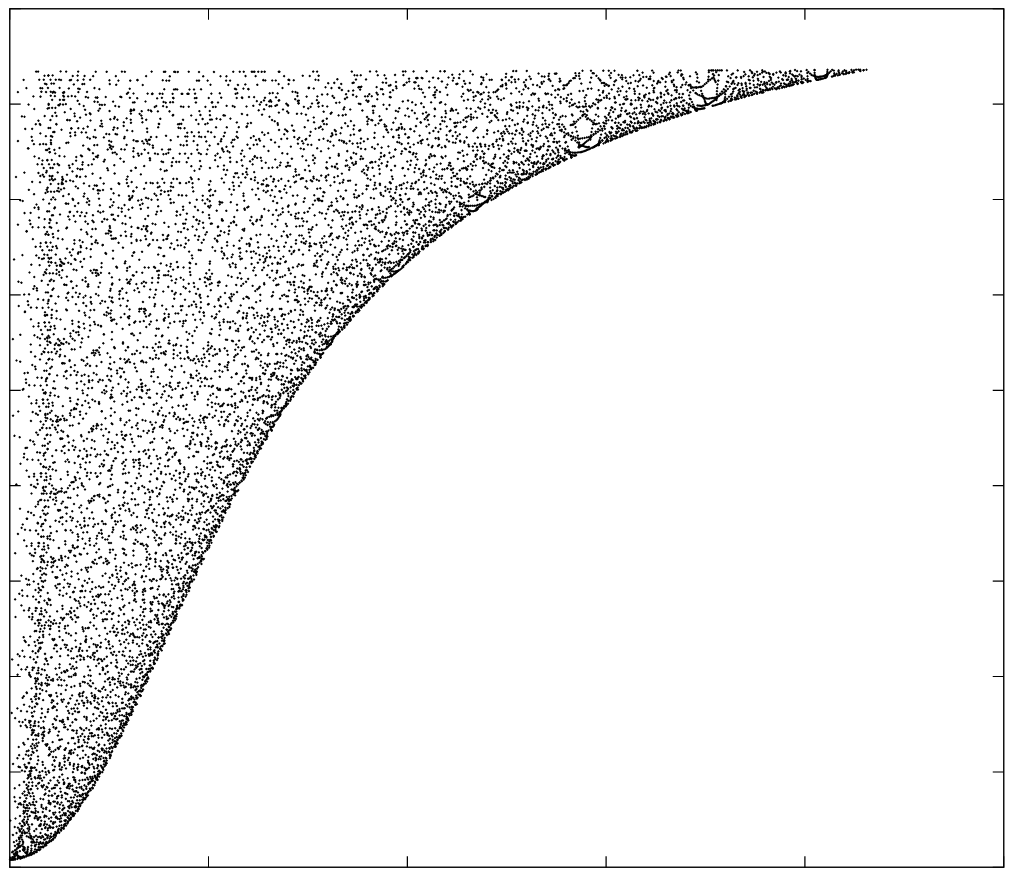}}\vspace*{-.16cm}
    \caption*{\hspace*{.7cm}\ft$t=0$}
  \end{subfigure}
  \begin{subfigure}[h]{0.49\textwidth}
    %\centering
    \resizebox{!}{.8\textwidth}{\input{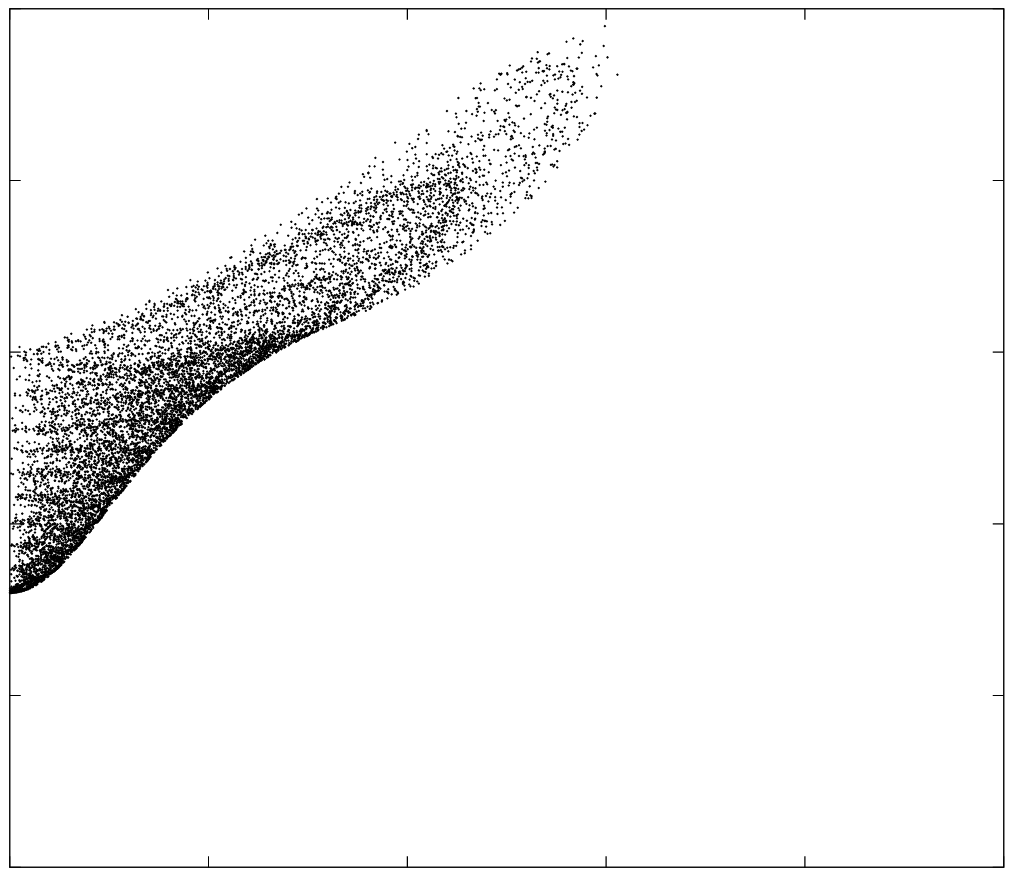}}\vspace*{-.16cm}
    \caption*{\hspace*{.85cm}\ft$t=15$}
  \end{subfigure}
	
  \vspace{.2cm}
	
  \begin{subfigure}[h]{0.49\textwidth}
    %\centering
    \resizebox{!}{.8\textwidth}{\input{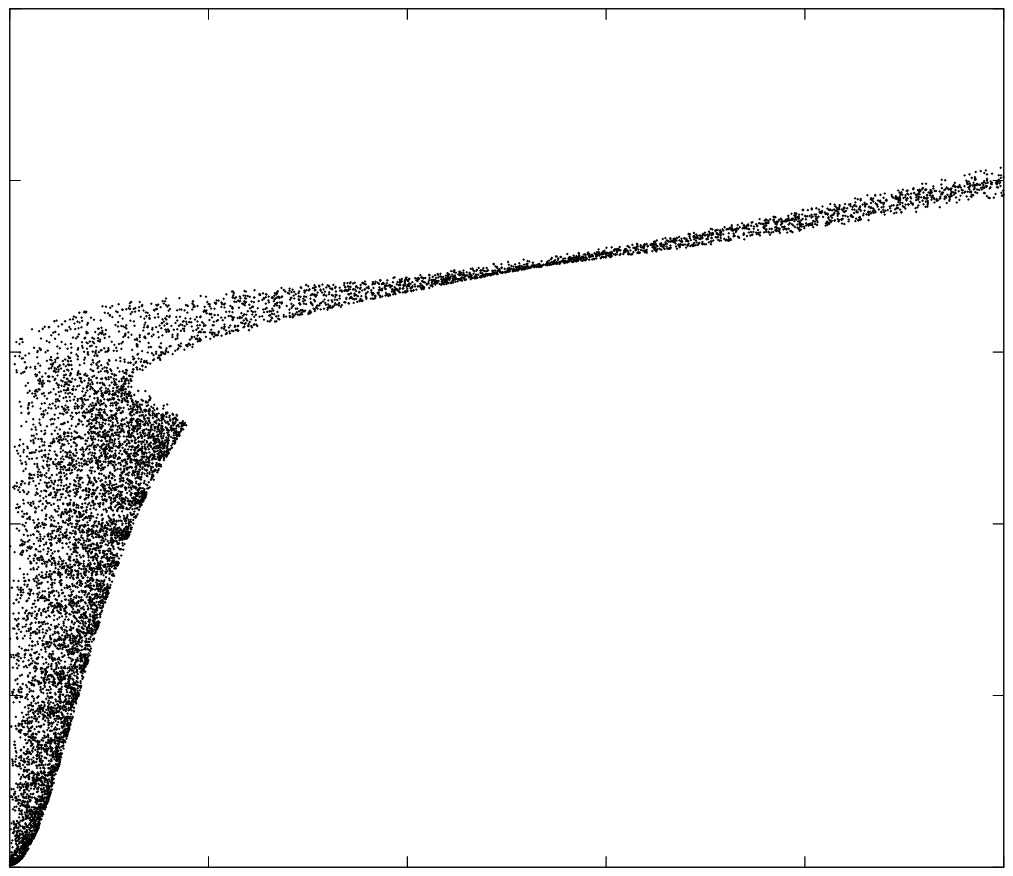}}\vspace*{-.16cm}
    \caption*{\hspace*{.85cm}\ft$t=30$}
  \end{subfigure}
  \begin{subfigure}[h]{0.49\textwidth}
    %\centering
    \resizebox{!}{.8\textwidth}{\input{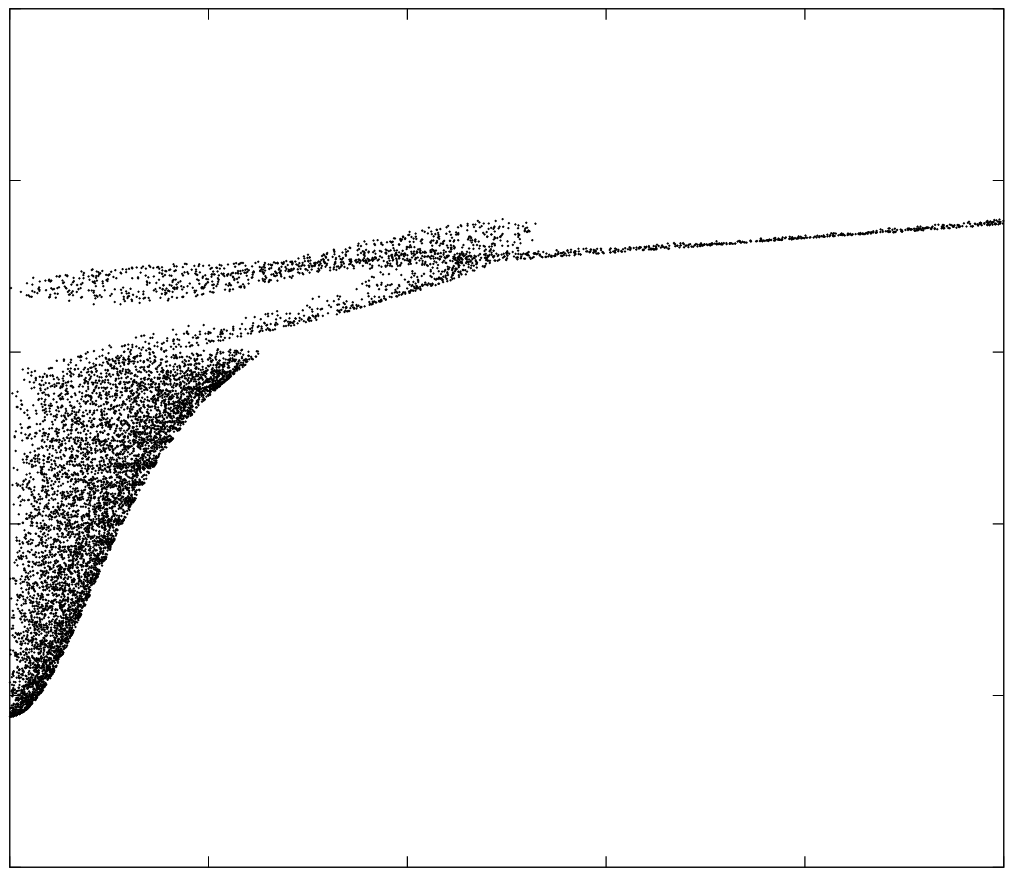}}\vspace*{-.16cm}
    \caption*{\hspace*{.85cm}\ft$t=45$}
  \end{subfigure}

  \vspace{.2cm}
	
  \begin{subfigure}[h]{0.49\textwidth}
    %\centering
    \resizebox{!}{.8\textwidth}{\input{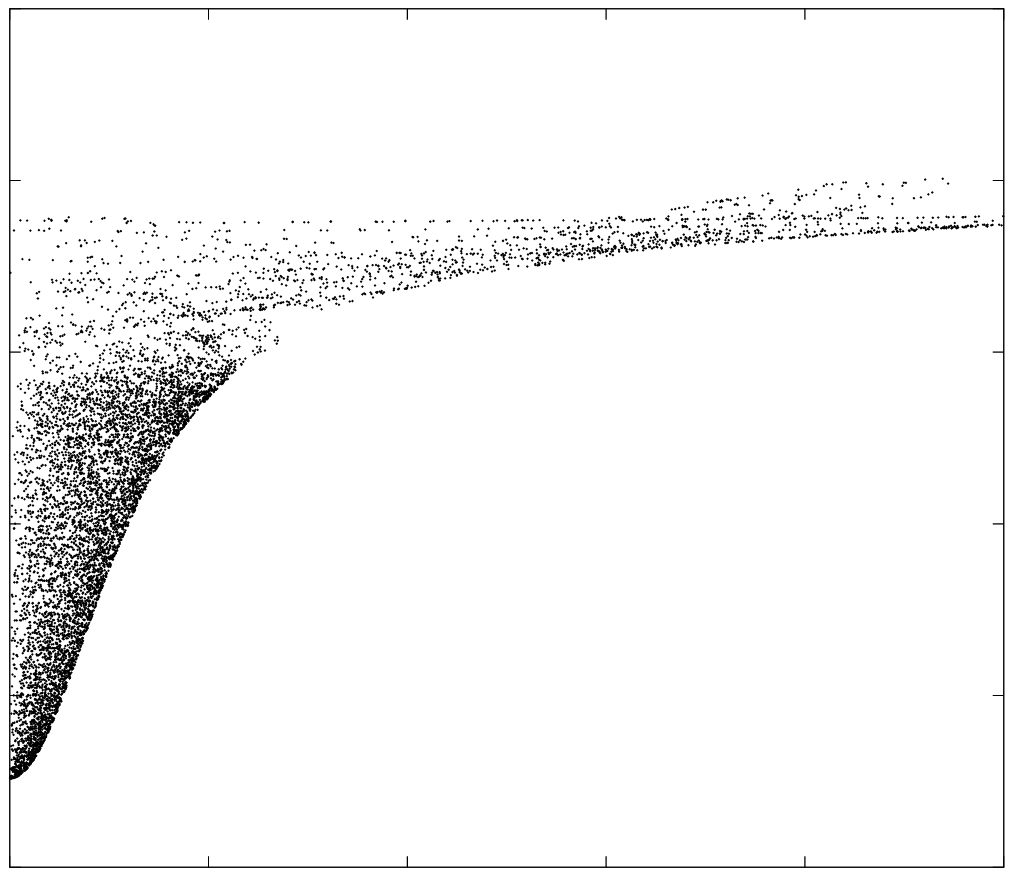}}\vspace*{-.16cm}
    \caption*{\hspace*{1.cm}\ft$t=120$}
  \end{subfigure}
  \begin{subfigure}[h]{0.49\textwidth}
    %\centering
    \resizebox{!}{.8\textwidth}{\input{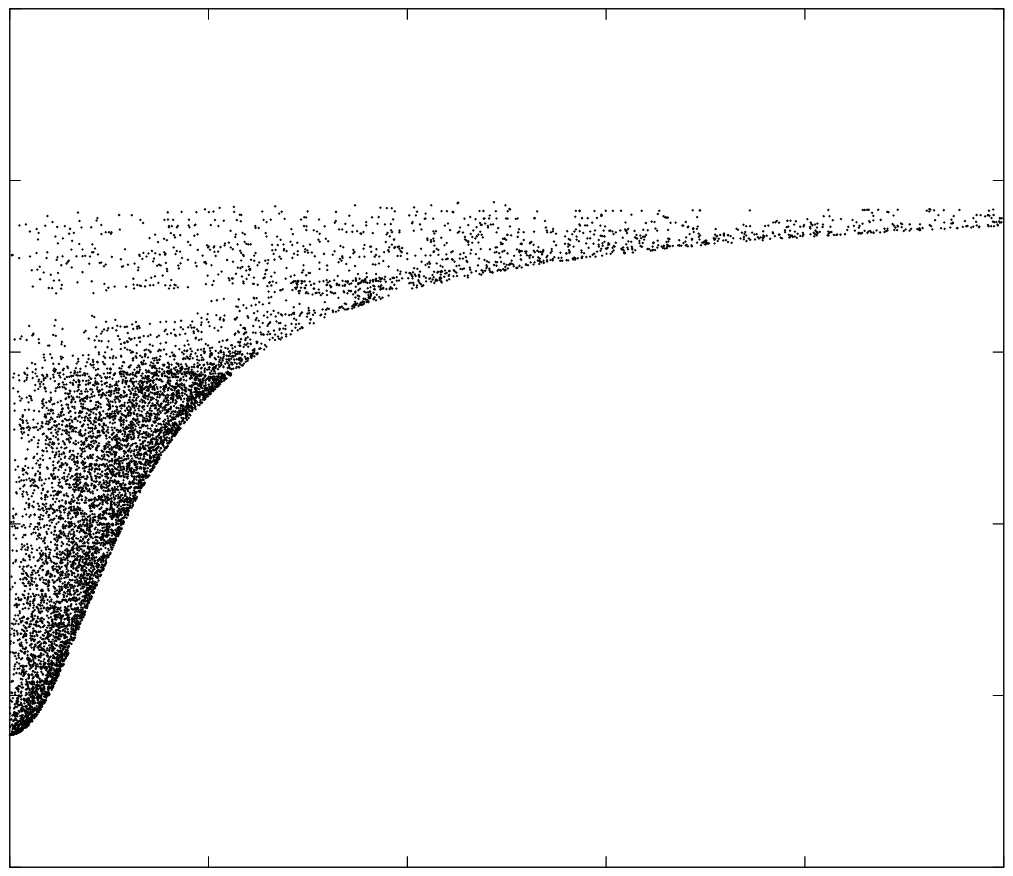}}\vspace*{-.16cm}
    \caption*{\hspace*{1.cm}\ft$t=180$}
  \end{subfigure}

  \caption{Numerical particles for the King model in Schwarzschild
    coordinates for $y_0 =0.6> y_{\mathrm{max}}$ perturbed by a dynamically
    accessible perturbation with direction $\gamma>0$. The movie corresponding to these snapshots is available at \cite{movies}.} 
  \label{img:strudel}
\end{figure}

On the particle level we observe that during the initial dispersion,
the particles spread in space. While some particles keep moving away from
the spatial origin manifesting the initial elevation, a considerable amount
returns after some time, which leads to a decrease of $e^{\mu(t,0)}$,
completing the first oscillation. This behavior of particles getting
expelled and returning repeats several times until the system eventually
oscillates around a seemingly unchanging configuration. In particular,
the oscillating behavior cannot only be seen in $e^{\mu(t,0)}$,
but also on the particle level as well as in all metric and matter
quantities. The oscillation of the mass density, i.e.,
the radial derivative of the Hawking mass, in the case of the King model
is depicted in Figure~\ref{img:rhofalsemap}.

Another illustration of the effect described above is given in
Figure~\ref{img:strudel} where we plot
the energy $E$ of individual particles against their distance from the
origin. Thereby, we can investigate trajectories of particles.
Note that the scales of both the energy $E$ and the radius $r$
change from the plot for $t = 0$ to the plot for $t =15$ and stay
the same for $t > 15$. Plotting these specific quantities allows
us to identify a portion of low energy particles that seems to form
some dense structure which is independent of the less dense particles
with higher energy. In addition, the dispersion of clusters of particles
can be observed in both Figure~\ref{img:rhofalsemap} and \ref{img:strudel}.

A possible interpretation of these observations is that
after perturbation a part of the solution corresponding to high
energy particles disperses while after some transition period
the remainder starts oscillating around
a new steady state. This new state is less relativistic than the
original one and seems to depend solely on
the original equilibrium, i.e., it is independent of numerical
parameters and the type and strength of the perturbation used to
perturb the original unstable steady state, at least for sufficiently
weak perturbations. 
In the context of dynamical systems such a new state is usually
a stable steady state of the system, and the above behavior is
reminiscent of a heteroclinic orbit, as the original steady state
after perturbation migrates to a different one. On the other hand,
it does not really seem to converge towards this new state,
as would be the case for a genuine heteroclinic orbit, so we
use this terminology very loosely.

Since the Einstein-Vlasov system possesses a plethora of steady states,
the explicit identification of the target state seems to be very delicate.
We developed a method which acts on the level of particles and basically
\enquote{deletes} all the \enquote{departed} particles from our
solution before fitting the remainder by some stable steady state.
However, a weak point of our procedure is that we have to decide
manually whether or not a particle belongs to the \enquote{remainder}.
In our simulations, we always made this decision based on the radius
or the particle energy. All this being said the observations with the
above procedure seem to support the  \enquote{heteroclinic orbit picture}.
Developing appropriate criteria for the
latter question would definitely open up new possibilities for
the search of these target states.
In passing we note that
Figure \ref{img:strudel} actually presents snapshots
from a movie which illustrates the above behavior and which,
together with similar simulations, can be viewed via the link \cite{movies}.

As can be seen in Figure~\ref{img:kingdis}, the difference between the
value of $e^{\mu(t,0)}$ around which the solution oscillates at later times
and the initial value $e^{\mu(0,0)}$ increases in $y_0$. In particular,
the initial elevation also increases in $y_0$. This is caused by a
relative increase of the number and mass of particles which initially
get expelled from the configuration as well as a decrease of the number
of particles which return. This difference seems to change
smoothly in $y_0$ which causes the initial elevation to be very
subtle for $y_0\approx y_{\mathrm{max}}$, cf.\ Figure~\ref{img:kingmaxdis},
consistent with the fact that no such elevation occurs in the stable regime.
Furthermore, the time-frame of the initial reimplosion of the solution
increases rapidly in $y_0$ when choosing $y_0$ large enough, which means
that the existence of fully dispersing solutions induced by large $y_0$
can not be ruled out numerically. 

\begin{figure}[h]
  \begin{center}
    \input{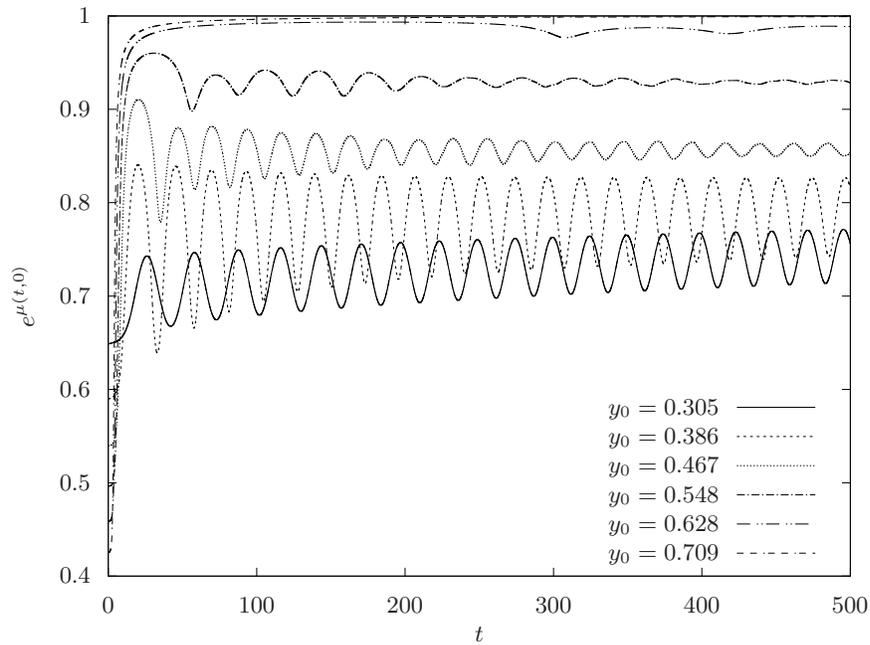}
  \end{center}
  \vspace*{-.7cm}
  \caption{The polytropic ansatz with
    $k=0.5$, $l=0.1$, $L_0=0.001$ in Schwarzschild coordinates with
    $y_0 > y_{\mathrm{max}}$ perturbed by a dynamically accessible
    perturbation with direction $\gamma>0$.}
  \label{img:polydis}
\end{figure}

It has been suggested in \cite{ShTe1985} that perturbations promoting
dispersion of certain isotropic steady states with negative binding energy
lead to fully dispersing solutions. In \cite{AnRe2006}, this suggestion has
been extended to non-isotropic steady states with an inner vacuum region,
i.e., $L_0>0$. However, our results disprove this conjecture in all
three coordinate systems under consideration. In the case of the King model,
the binding energy of a steady state is negative for $y_0>0.784$ which means
that Figure~\ref{img:kingdis} clearly shows the reimplosion and oscillation
of solutions emerging from weakly perturbed steady states with negative
binding energy. The same phenomenon can also be observed for other isotropic
and non-isotropic models, at least if $L_0$ is not too large.
In the case of a polytropic ansatz with $k=0.5$, $l=0.1$, $L_0=0.001$
this is depicted in Figure~\ref{img:polydis}. 

Notice that $L_0$ is closely connected to the size of the inner
vacuum region of steady states. Increasing this parameter in the
ansatz function seems to slow down all occurring effects, in particular
causing the reimplosion time to increase. This effect is the main reason
why we choose $L_0$ rather small. 
In fact, when setting $L_0$ significantly larger, we cannot decide
whether some steady states with $y_{\mathrm{max}}<y_0<y_{\mathrm{zero}}$
reimplode or fully disperse. A similar effect can be observed for large $k$.
Furthermore, large $y_0$ pose a problem when examining dispersing steady
states for every model since we cannot numerically distinguish solutions
with large reimplosion times from fully dispersing ones.   
It therefore remains an open question whether or not there exist
fully dispersing solutions emerging from weakly perturbed unstable
steady states and if this behavior is connected to the binding energy
of the original steady states in some way.

\section{Discussion of the numerics}\label{sc:errors}
\setcounter{equation}{0}

We conclude our paper with an overview of the parameter settings which
influence the numerical accuracy. Compared to former numerical projects
in this field of research, we benefit from more computational power and
from a parallelized code. This allows us to perform computations with
a large number of numerical particles within a reasonable computation time.
The actual computation time further depends on the discretization of the
time variable $t$ and the spatial variable $r$. We typically use $\Delta r$
and $\Delta t$ in the order of magnitude of $10^{-4}$. Moreover,
we ensure that at least $15$ million numerical particles are used
for our computations by choosing the number of steps $N_u$ and $N_{\psi}$
suitably after fixing $\Delta r$ and $\Delta t$. In order to guarantee that
perturbations resulting of errors due to our initialization are small
compared to our applied perturbation and that the applied perturbation is
reasonably small we choose $T_{pert}=0.5$ and $\epsilon_{pert}$
of the order $10^{-4}$ and determine a suitable $\gamma$ as described
in Section~\ref{sc:numeric}.

To monitor the validity of our simulation, we keep track of the ADM mass
$M(t)$ and the analytical number of particles $N(t)$ which are conserved
quantities along solutions of the Einstein-Vlasov system.
We define the relative errors as 
\[
e_M(t) = \frac{|{M(t)-M}|}{M}, \quad e_N(t) = \frac{|N(t)-N|}{N}
\]
where $M=M(T_{pert})$ and $N=N(T_{pert})$ are given by the perturbed steady
state, i.e., the evolved state at $t=T_{pert}$. In the case of oscillating
or heteroclinic solutions the errors stay very small---of the order $10^{-4}$
until $t= 100$---when choosing the numerical parameters as
mentioned above. When considering the formation of a trapped surface and the
formation of a black hole, it turns out that the errors become larger
and---not surprisingly---the
simulation eventually breaks down.

A further test of our codes is to evolve an unperturbed steady state.
When choosing a stable steady state it is tracked faithfully
for very long times. Obviously, for an unstable steady state the errors
due to the initialization can eventually cause a deviation from the
steady state.
In conclusion, it seems fair to say that our simulations provide
conclusive results at least if we are not considering the long
time behavior of collapsing solutions after trapped surfaces have formed.

\bigskip

%\noindent
%    {\bf Acknowledgments.}
%    This paper originates from a practical training unit in the master
%    program in mathematics at the University of Bayreuth, which one of the
%    authors (the teacher) offered in 2019 to his coauthors (the students),
%    and he/they/we learned a lot.
%    

\end{document}